\pdfoutput=1
\documentclass[fleqn,10pt]{wlscirep}
\usepackage[utf8]{inputenc}
\usepackage[T1]{fontenc}
\usepackage{graphicx}
\usepackage{subcaption}
\usepackage{geometry}
\usepackage{xcolor}
\usepackage{setspace}
\usepackage{url}

\title{Neighborhood Deprivation Shapes Post-Flood Population Losses in Pune and Bandung}

\author[1,2,+]{Sukie Xiuqi Yang}
\author[3,4,+]{Guangyuan Weng}
\author[5]{Shiyue Cui}
\author[6]{Zhicong Chen}
\author[7,8,*]{Catherine Chen}

\affil[1]{Urban Governance and Design Thrust, The Hong Kong University of Science and Technology (Guangzhou), Guangzhou, China}
\affil[2]{Carbon Neutrality and Climate Change Thrust, The Hong Kong University of Science and Technology (Guangzhou), Guangzhou, China}
\affil[3]{Khoury College of Computer Sciences, Northeastern University, Boston, MA 02115, USA}
\affil[4]{Network Science Institute, Northeastern University, Boston, MA 02115, USA}
\affil[5]{Department of Sociology and Criminology, University at Buffalo-SUNY, Buffalo, NY 14260, USA}
\affil[6]{Department of Communications and New Media, National University of Singapore, Singapore, Singapore}
\affil[7]{Manship School of Mass Communication, Louisiana State University, Baton Rouge, LA 70803, USA}
\affil[8]{Department of Political Science, Louisiana State University, Baton Rouge, LA 70803, USA}
\affil[*]{Corresponding author: catherinechen@lsu.edu}
\affil[+]{These authors contributed equally to this work.}

\makeatletter
\renewcommand{\@maketitle}{%
  \vspace*{-20pt}%
  {\raggedright\sffamily\bfseries\fontsize{20}{25}\selectfont\@title\par}%
  \vspace{18pt}%
  {\raggedright\setstretch{1.15}\@author\par}%
}
\makeatother

\begin{abstract}
Disaster research in high-income settings shows that better-resourced communities see population inflows and faster recovery, while disadvantaged communities suffer prolonged population loss. Whether this pattern holds in rapidly urbanizing cities of the Global South remains unclear. Here we show, using mobile phone data from Pune, India, and Bandung, Indonesia, how neighborhood-level changes in phone presence, a proxy for short-term population movement, relate to flooding and neighborhood deprivation. We combine satellite-derived flood maps with phone records to compare flooded and non-flooded areas over the 30 days around each flood. In both cities, less deprived neighborhoods see phone presence rise after flooding, while more deprived neighborhoods see it fall, consistent with \textcolor{black}{resource-seeking interpretations} documented in North American and European disaster research. The cities differed, however, in how \textcolor{black}{deprivation relates to flood exposure: in Pune, more deprived flooded neighborhoods showed suggestive evidence of disproportionately larger decreases in detected device presence than similarly deprived non-flooded neighborhoods; no comparable interaction was observed in Bandung.} Economic considerations are associated with \textcolor{black}{post-flood device-presence changes} in South and Southeast Asian cities in ways that parallel high-income contexts, though local context adds variation.

\end{abstract}

\begin{document}
\pagestyle{empty} % Main manuscript without page numbers.
\maketitle
\thispagestyle{empty}
\clearpage
\section*{Abstract}
{\sffamily\small\noindent\theabstract\par}

\vspace{1em}

\newpage
\setcounter{page}{1}
\section*{Introduction}

Climate change has substantially increased the frequency and intensity of floods globally, reshaping patterns of human mobility and settlement \cite{ipcc_evidence_2023}. While overall flood management capacity has improved in parts of the world, disaster-related losses continue to rise in many low-income and rapidly urbanizing regions, reflecting widening inequalities in risk exposure and recovery capacity \cite{formetta_empirical_2019}. In contrast to many cities in the Global North, where extensive drainage networks, flood defenses and post-disaster recovery programs mitigate long-term displacement, many regions in the Global South face inadequate drainage systems and uneven infrastructure investment, making them disproportionately exposed to prolonged disruption and loss of livelihood \cite{yamamoto_human_2017}. For example, flooding associated with Cyclone Amphan displaced millions of people in India and Bangladesh in 2020, stretching beyond physical destruction to short-term evacuation, temporary relocation, and longer-term changes in residential and commuting patterns \cite{hassan_quantitative_2020}. Many households were forced to seek refuge on embankments and in cyclone shelters, amid saline water inundation, vegetation loss, and freshwater scarcity, reflecting limited recovery capacity and fragile livelihood systems \cite{rashid2023consequences}. Understanding mobility in these Global South contexts therefore has important implications for immediate humanitarian response and long-term urban planning \cite{thomalla_reducing_2006, birkmann2013framing, graveline2022disaster}. More broadly, it also contributes to theoretical debates on disaster resilience and human adaptation by foregrounding how structural inequalities shape mobility responses to environmental shocks \cite{gioli_human_2016,guadagno_human_2017,oakes_future_2023}.

Much of the foundational theory around disaster mobility has emerged from Western contexts, particularly the United States, where formal infrastructure, insurance systems, and state-led evacuation frameworks shape how mobility is organized and understood \cite{priovashini_bibliometric_2022, tarun2025rethinking, milan2021climate}. Early sociological work emphasized socioeconomic status (SES) as a key determinant of recovery capabilities, highlighting how low-income individuals face greater barriers to evacuate and recover \cite{fothergill2004poverty}. This literature expanded through major disaster events, most notably Hurricane Katrina, showing how race, class, and gender intersected to shape post-disaster trajectories \cite{groen2008hurricane, elliott2006race, fussell_measuring_2014}. More recent studies using large-scale mobility data continue to highlight socioeconomic disparities, while also introducing more granular insights into spatial and temporal mobility behaviors \cite{hong_measuring_2021, deng_examining_2021, smiley2022social, weng_beyond_2025}. Hong et al.\cite{hong_measuring_2021} analyzed over 800,000 devices during Hurricane Harvey and found that low-income, minority neighborhoods faced the most prolonged mobility disruptions, while Li et al.\cite{li_mobility_2024} shows that minority and low-income populations spend significantly more time in flood-prone areas due to daily movement, even though their homes are less exposed. These studies extend existing research, contributing to an increasingly coherent framework for analyzing post-disaster mobility in high-income contexts.

While U.S. disaster research consistently documents socioeconomic gradients in post-disaster mobility, with more advantaged neighborhoods recovering more quickly and attracting displaced populations, cities in South and Southeast Asia and Latin America operate under markedly different structural conditions \cite{nixon2011slow, bankoff2003cultures, gaillard2010vulnerability}. These conditions are characterized by widespread informal housing, fragmented land tenure, uneven infrastructure provision, and limited access to formal relocation and compensation mechanisms \cite{hussainzad2024climate, yamamoto_human_2017, de2024mapping}. These features do not necessarily eliminate socioeconomic differentiation, but call into question whether well-documented mobility gradients observed in high-income settings operate in similar ways under contrasting institutional environments. Empirical research in Nepal, Indonesia, and the Philippines shows that displacement decisions may be influenced by terrain, traditional housing forms, kinship ties, and ancestral land access \cite{lam_cultural_2011, lam_displacement_2013,shalih_toward_2020, galura_social_2018}, as well as collective hazard knowledge and locally adapted mobility practices \cite{marino_long_2012, dede2024disaster}. Examples include communal evacuation strategies and resilient housing types like rumah panggung (stilt houses) in Indonesia \cite{dede2024disaster, yani_rumah_2016, herlina_local_2025}. Taken together, this literature highlights substantial contextual variation in disaster-related mobility, while underscoring the lack of systematic evidence on whether well-documented patterns from high-income settings extend to rapidly urbanizing low and mid-income contexts.

\begin{figure}[htbp]
    \centering
    \includegraphics[width=\textwidth,trim=12bp 12bp 12bp 12bp,clip]{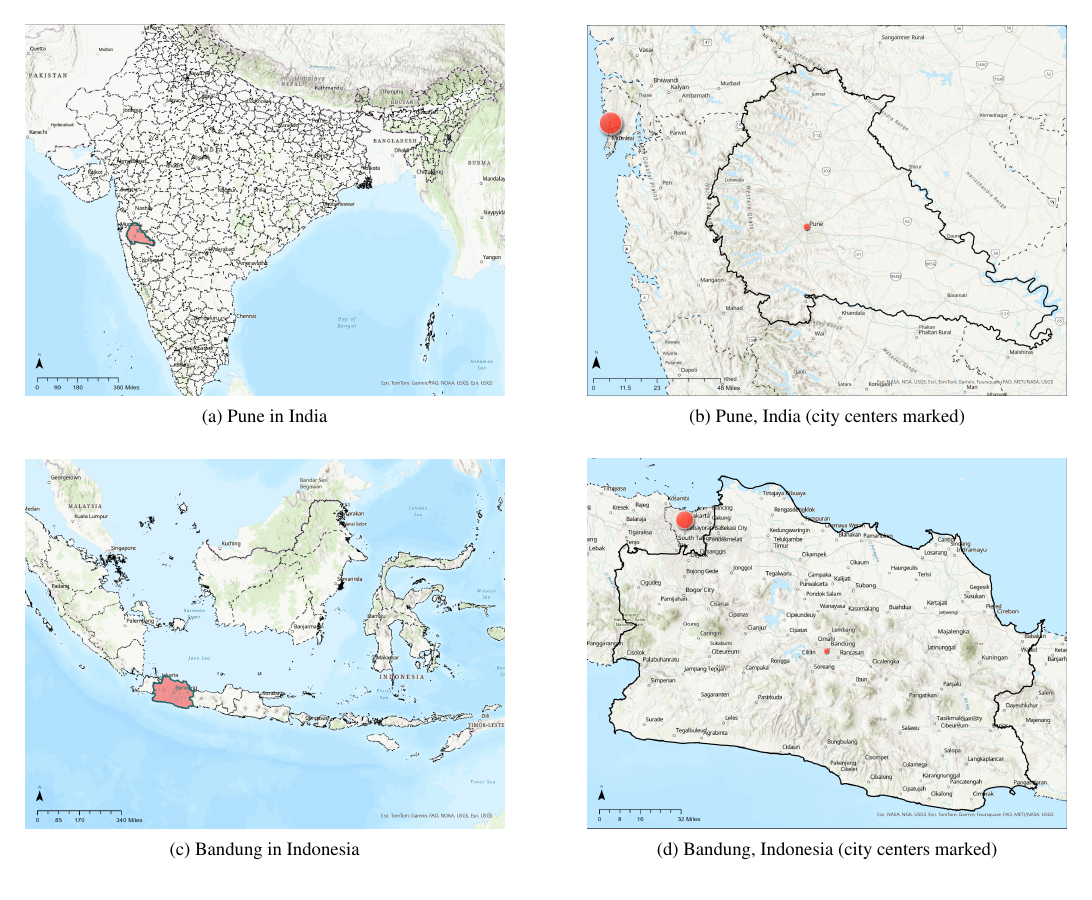}
    \caption{Geographical context of flood events in Pune, India and Bandung, Indonesia}
    \label{fig:geographical_context}
\end{figure}

\textcolor{black}{Building on this literature, we focus on rapidly urbanizing, disaster-prone contexts in Monsoon Asia that remain underrepresented in large-scale mobility research. India and Indonesia provide two such cases, given their diverse geography, large populations, and pronounced socioeconomic disparities  \cite{milan2021climate, dhiman_flood_2018, abdillah2025adapting}. We select city cases within these contexts based on three dimensions: data availability, metropolitan population scale, and flood characteristics \cite{gerring2006case}. We linked device-based mobile phone data from the NetMob 2024 dataset \cite{zhang_netmob2024_2024} as proxies for neighborhood-level mobility, with EM-DAT disaster records for 2019 and 2020 \cite{delforge2023dat}, and we selected Pune, India and Bandung, Indonesia as comparative cases since both meet these criteria with consistent daily data availability, comparable urban scale, and \textcolor{black}{major} flood events during this period. Pune, with approximately 7.4 million residents, serves as Maharashtra's second-largest city and a major industrial, educational, and military center \cite{un_world_population_prospects_pune_2024}, while Bandung, home to roughly 8.2 million people in its metropolitan area, functions as West Java's provincial capital and a major manufacturing, educational, and cultural hub \cite{badan_pusat_statistik_bps_kota_2020}(See Figure \ref{fig:geographical_context}). Both cities operate as critical transportation nodes connecting rural areas with national networks, house prestigious universities and governmental institutions, experience rapid urbanization pressures, and face recurring flood vulnerability due to their monsoon-affected locations \cite{butsch_growing_2017,tarigan_bandung_2016}. Both cities experienced Class 1 monsoon-driven events affecting \textcolor{black}{broad urban areas}, with at least 30 days of continuous pre-event device counts above the 10-device privacy threshold for at least 150 grid cells. These shared characteristics make them suitable comparative cases for examining disaster mobility patterns in secondary cities of the Global South. We do not treat these cases as strictly equivalent, but as sufficiently comparable for identifying patterned variation. Other potential study sites in the dataset lacked either comparable disaster impacts or sufficient continuous data coverage, making the Pune-Bandung comparison particularly valuable for understanding regional patterns of disaster mobility in Monsoon Asia. This study design allows us to explore how inequality and disaster risk interact in contexts where disaster governance, market access, and social protection are more unevenly developed than in major metropolitan centers.}

This study addresses two research questions. First, what distinct post-flood mobility trajectories can be identified across neighborhoods using device-based mobility data in Pune and Bandung? Second, how are these trajectories associated with neighborhood deprivation and flood exposure under conditions of limited institutional capacity? Leveraging mobile phone data at a spatial resolution of 4.9km-by-4.9km, we use an agglomerative hierarchical clustering algorithm to identify recurring mobility patterns and explore regional variations. Our analysis reveals a consistent bifurcation in post-flood mobility across Pune and Bandung, with neighborhoods clustering into \textcolor{black}{trajectories characterized by increases and decreases in detected device presence}. In both cities, areas experiencing \textcolor{black}{increases in detected device presence} are characterized by lower deprivation, stronger built infrastructure, and higher nighttime light intensity, indicating systematic socioeconomic stratification in post-disaster mobility. Logistic regression models further \textcolor{black}{corroborate} these associations and demonstrate that deprivation \textcolor{black}{is significantly associated with} cluster membership in both settings. While this socioeconomic gradient is present in both cities, the association between flood exposure and \textcolor{black}{decreases in detected device presence} was \textcolor{black}{moderated} by neighborhood deprivation only in Pune, in which more deprived neighborhoods that flooded experienced disproportionately larger \textcolor{black}{decreases in detected device presence} than comparably deprived neighborhoods outside the flood zone. \textcolor{black}{We treat this interaction as suggestive: nearly all Pune cells fall inside the flood polygon, so it is identified off a narrow band of variation. No comparable interaction was observed in Bandung.} Together, these findings indicate that post-flood mobility was associated with neighborhood socioeconomic conditions in both cities, while local urban and environmental contexts accounted for additional variation between them. These findings expand established theoretical frameworks developed in the context of North America and Europe, highlighting socioeconomic factors in disaster mobility with diverse social contexts. It reveals that socioeconomic gradients in post-disaster mobility persist even under conditions of institutional scarcity and infrastructural constraint. Where formal disaster governance and social protection are limited, economic and spatial advantage appear associated with differential mobility outcomes.

\section*{Results}

\subsection*{Post-flood \textcolor{black}{device-presence changes} in Pune and Bandung}

\begin{figure}[htbp]
    \centering
    \includegraphics[width=\textwidth,trim=12bp 12bp 12bp 12bp,clip]{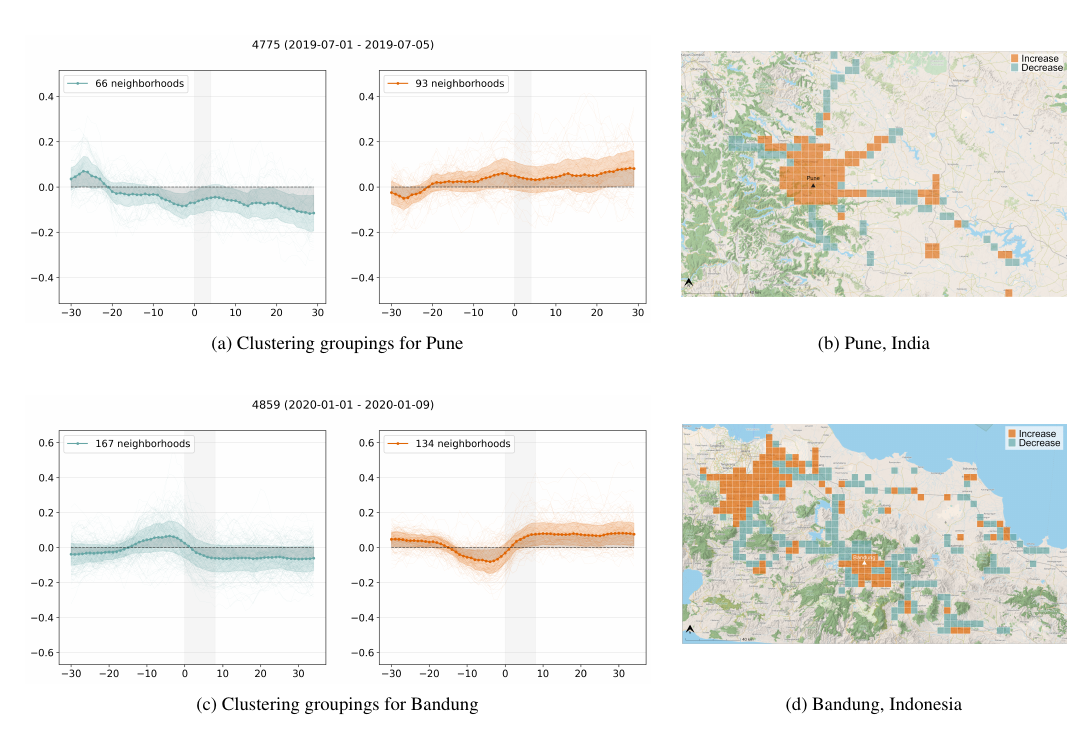}
    \caption{Community clustering results and the corresponding mapping}
    \label{fig:all_figures}
\end{figure}

Our analysis draws on two study sites with comparable data quality and flood exposure. In Pune (Figures \hyperref[fig:all_figures]{2a} and \hyperref[fig:all_figures]{2b}), we analyzed 159 grid cells across the Pune district that had sufficiently complete mobility trajectories (no more than three consecutive days of missing data). The July 2019 flood event (lasting from July 1 to July 5) \textcolor{black}{intersected approximately 90\% of these cells based on Dartmouth Flood Observatory (DFO) flood-event polygons (mean cell-area overlap: 87\%; see Table~\ref{cluster_characteristics})}. In Bandung (Figures \hyperref[fig:all_figures]{2c} and \hyperref[fig:all_figures]{2d}), we analyzed 301 grid cells across West Java province with sufficiently complete mobility trajectories. This pre-COVID, January 2020 flood event (lasting from January 1 to January 9) \textcolor{black}{intersected approximately 74\% of these cells based on DFO polygons (mean cell-area overlap: 69\%; see Table~\ref{cluster_characteristics})}.

Agglomerative hierarchical clustering of neighborhood mobility trajectories yielded a two-cluster solution in both cities. Silhouette scores, which measure within-cluster cohesion relative to between-cluster separation, peaked at k=2 for Pune (0.35) and Bandung (0.30), declining with additional clusters. Dendrograms and Elbow plots also \textcolor{black}{suggested} this optimal cluster number (see Appendix Figure A1). These two clusters correspond to neighborhoods that experienced \textcolor{black}{increases versus decreases in detected device presence, used here as a proxy for short-term population presence,} following a flood event (Figure \ref{fig:all_figures}). Orange indicates \textcolor{black}{increases in detected device presence} and teal indicates \textcolor{black}{decreases in detected device presence.} In the time-series panels, day 0 marks the onset of each flood event, and the shaded region denotes the flood duration as recorded in NASA flood archives.

In Pune, neighborhoods with \textcolor{black}{decreases in detected device presence} ($n=66$) were distributed across peripheral areas, particularly in the northern and western portions of the study region. Neighborhoods with \textcolor{black}{increases in detected device presence} ($n=93$) were located primarily along the main urban corridor running through Pune and Pimpri-Chinchwad, extending into the eastern districts.

In Bandung, neighborhoods with \textcolor{black}{decreases in detected device presence} ($n=167$) concentrated in transitional zones between the urban core and the mountainous periphery. Neighborhoods with \textcolor{black}{increases in detected device presence} ($n=134$) appeared both within the Bandung metropolitan core and scattered across the southern and eastern sectors of West Java.

The temporal profiles (Figures \hyperref[fig:all_figures]{2a} and \hyperref[fig:all_figures]{2c}) illustrate both the magnitude and persistence of these \textcolor{black}{device-presence} shifts, with baseline defined as the mean device count over the 30 days preceding flood onset. In Pune, the decrease cluster hovered roughly 5–10\% below baseline following the flood, while the increase cluster rose gradually, reaching approximately 10\% above baseline by day 30. Bandung displayed more modest shifts: the decrease cluster settled around 5\% below baseline, and the increase cluster climbed to approximately 5\% above baseline in the weeks after the flood.

\subsection*{Socioeconomic characteristics of mobility clusters}

\begin{figure}[htbp]
    \centering
    \includegraphics[width=0.92\textwidth,trim=12bp 12bp 12bp 12bp,clip]{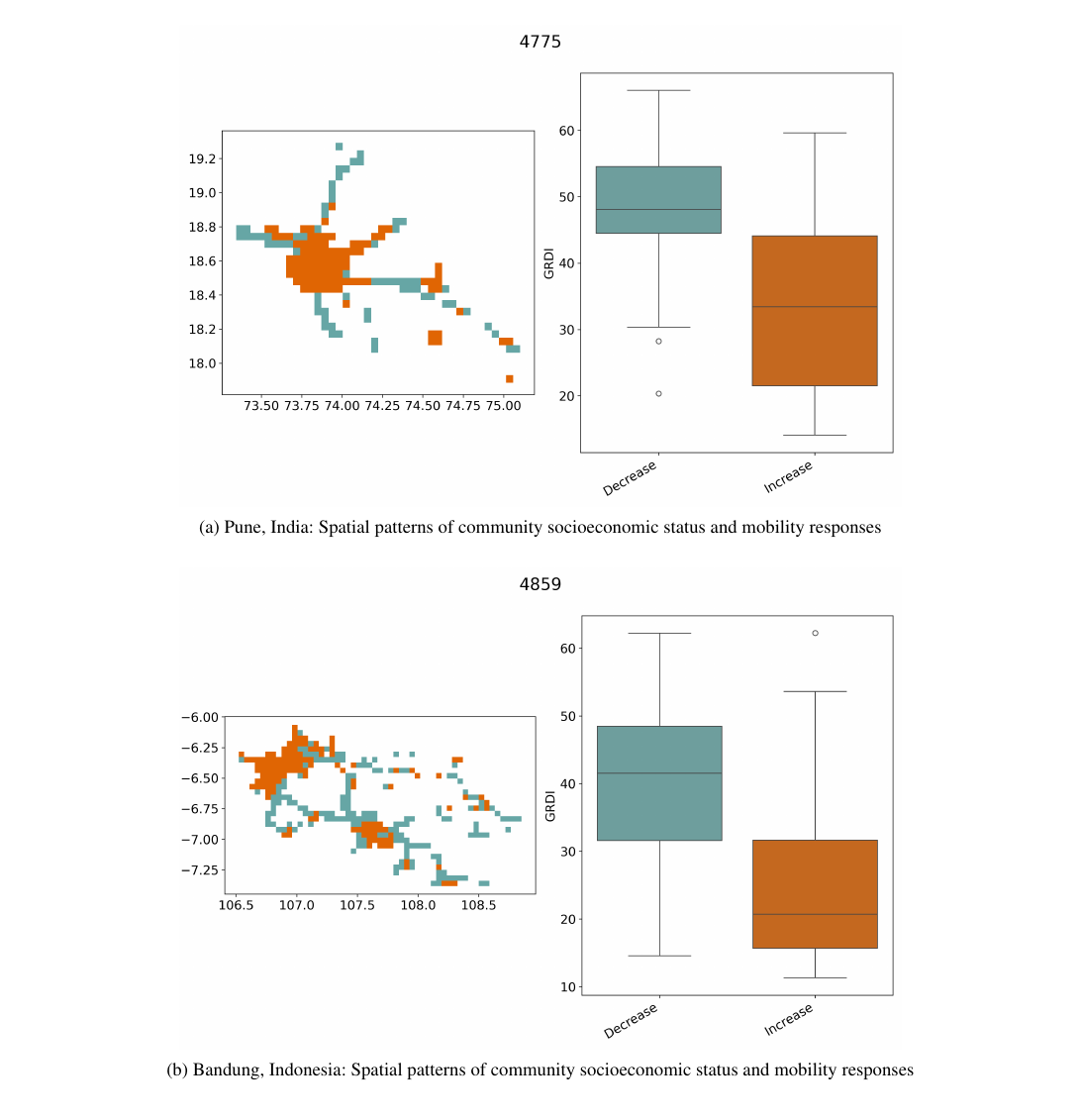}
    \caption{Mapping community deprivation levels and mobility patterns in response to flooding events.}
    \label{fig:combined_figures}
\end{figure}

Neighborhood socioeconomic status and post-flood mobility showed a positive association in both cities: areas with lower deprivation levels were more likely to experience \textcolor{black}{increases in detected device presence} (Figure \ref{fig:combined_figures}, Table \ref{cluster_characteristics}). In Pune and Bandung alike, neighborhoods that \textcolor{black}{experienced increases in detected device presence} during flooding had lower deprivation scores than neighborhoods that \textcolor{black}{showed decreases in detected device presence}. The boxplot distributions in Figure \ref{fig:combined_figures} display this gradient: in Pune, the decrease cluster shows a median GRDI around 48, while the increase cluster centers near 33. In Bandung, the decrease cluster has a median GRDI of around 40, compared with approximately 22 for the increase cluster.

Table \ref{cluster_characteristics} (top panel) presents the cluster-level characteristics for Pune. Neighborhoods experiencing \textcolor{black}{decreases in detected device presence} had a mean deprivation index of 48.48, whereas those with \textcolor{black}{increases in detected device presence} averaged 33.42. Built-up infrastructure was more developed in areas \textcolor{black}{with increases in detected device presence} (65.79) than in areas \textcolor{black}{with decreases in detected device presence} (41.80). Nighttime light intensity---a proxy for economic activity---was nearly twice as high in the increase cluster (65.32) as in the decrease cluster (33.65). Nighttime light change over time, which captures recent economic growth trajectories, was also higher in the increase cluster (72.48) than in the decrease cluster (52.76). Elevation differed modestly between the two clusters, with the decrease cluster averaging 616.4 meters and the increase cluster 586.9 meters.

Bandung exhibited a similar socioeconomic gradient (Table \ref{cluster_characteristics}, bottom panel). Neighborhoods with \textcolor{black}{increases in detected device presence} had a mean deprivation index of 24.43, compared with 39.36 for neighborhoods with \textcolor{black}{decreases in detected device presence}. Built-up infrastructure was higher in the increase cluster (20.38) than in the decrease cluster (12.26). Nighttime light intensity showed a similar difference: 59.63 in areas \textcolor{black}{with increases in detected device presence} versus 34.95 in areas \textcolor{black}{with decreases in detected device presence}. Nighttime light change over time followed the same direction (46.48 versus 24.31). Elevation diverged more sharply in Bandung than in Pune: the decrease cluster averaged 424.7 meters, while the increase cluster averaged 224.8---indicating that \textcolor{black}{increases in detected device presence} occurred at substantially lower elevations.

Both areas thus exhibited a common pattern: \textcolor{black}{neighborhoods with increases in detected device presence tended to have} stronger economic infrastructure, greater nighttime light intensity, and lower deprivation. This finding is consistent with resource-seeking \textcolor{black}{interpretations} documented in disaster mobility research from high-income settings, where wealthier areas with better services \textcolor{black}{show increases in detected device presence}. The consistency of this pattern across two cities in different countries suggests that economic considerations \textcolor{black}{correlate with} post-flood \textcolor{black}{device-presence change} in South and Southeast Asian contexts in ways that parallel high-income settings.

\subsection*{Socioeconomic and geographic correlates of post-disaster \textcolor{black}{device-presence change}}

To examine neighborhood characteristics associated with \textcolor{black}{device-presence change}, we estimated binomial logistic regression models for both cities (Table \ref{tab:logit_combined}). The dependent variable is a dichotomous indicator of cluster membership derived from the hierarchical clustering analysis described above. Independent variables include the Global Relative Deprivation Index (GRDI), flood exposure status, elevation terms, and an interaction between GRDI and flood exposure.

In Pune (Models 1 and 2), the GRDI coefficient is negative and statistically significant in the baseline model \textcolor{black}{(coefficient = $-0.116$, $p < 0.001$)}. \textcolor{black}{For each one-unit increase in the deprivation index, the odds of a neighborhood falling into \textcolor{black}{the cluster with increases in detected device presence (rather than the cluster with decreases in detected device presence)} declined by approximately eleven percent.} The main effect of flood overlap is small and not statistically distinguishable from zero in Model 1. In Model 2, once the interaction is included, the flood-overlap coefficient becomes positive and significant at conventional levels (coefficient = $0.069$, $p < 0.05$), and the GRDI $\times$ flood-overlap interaction is negative and marginally significant (coefficient = $-0.0013$, $p < 0.1$). The interaction coefficient is small per unit because flood overlap is measured on a 0--100 scale, \textcolor{black}{but the implied change across that range is non-trivial}: at zero overlap the implied GRDI slope is approximately $-0.004$ and indistinguishable from zero, while at 100 percent overlap the implied slope steepens to roughly $-0.130$. In other words, deprivation predicted \textcolor{black}{decreases in detected device presence} most strongly in cells that were fully inside the flood polygon, and barely at all in cells outside it. We read this as consistent with a differential-vulnerability pattern: \textcolor{black}{where flood exposure is high, deprivation is more strongly associated with whether a cell experiences increases or decreases in detected device presence} than it does in unaffected cells \cite{ronco2023exploring, cutter2003social}. \textcolor{black}{Because roughly 90 percent of Pune's analytic cells fall inside the flood polygon, this interaction rests on a relatively narrow band of within-city variation along the polygon margin and should be read as suggestive rather than definitive.}

In Bandung (Models 3 and 4), \textcolor{black}{the GRDI coefficient is again negative and statistically significant in the baseline model (coefficient = $-0.104$, $p < 0.001$), with the odds of \textcolor{black}{increases in detected device presence} falling by roughly ten percent per unit increase in deprivation. The flood-overlap main effect is small and not statistically distinguishable from zero in either Model 3 (coefficient = $-0.007$, $p > 0.10$) or Model 4 (coefficient = $0.009$, $p > 0.10$). The GRDI $\times$ flood-overlap interaction is essentially zero. This \textcolor{black}{suggests} that neighborhood deprivation predicts cluster membership about equally inside and outside the flood polygon.}

\subsection*{Common mobility patterns across Pune and Bandung}

Both cities exhibited a similar pattern: neighborhoods with lower deprivation and stronger economic activity \textcolor{black}{tended to experience increases in detected device presence}, while more deprived neighborhoods tended to \textcolor{black}{experience decreases}. This consistency across two cities in different countries suggests that the importance of economic factors in post-flood \textcolor{black}{device-presence change} may extend beyond the high-income contexts where such gradients have been most thoroughly documented. \textcolor{black}{Neighborhoods with increases in detected device presence tended to have lower deprivation}---a pattern \textcolor{black}{consistent with} resource-seeking \textcolor{black}{interpretations} documented in disaster research from North America and Europe.

In both cities, \textcolor{black}{neighborhood deprivation was robustly associated with the probability of \textcolor{black}{decreases in detected device presence}. The GRDI coefficient was negative in the baseline specifications (Pune: $\beta \approx -0.12$; Bandung: $\beta \approx -0.10$; both $p < 0.001$; Table~\ref{tab:logit_combined}, Models 1 and 3): a one-unit increase in the GRDI lowered the odds of belonging to \textcolor{black}{the cluster with increases in detected device presence} by roughly 10 to 11 percent in each area.} \textcolor{black}{The GRDI $\times$ flood-overlap interaction was small and negative in both cities (Pune: $\beta = -0.0013$, $p < 0.1$; Bandung: $\beta = -0.0001$, n.s.), suggesting that the deprivation gradient steepens slightly with flood exposure. We treat both interactions as suggestive rather than definitive and return to them in the Discussion.}

At least in the short term and at the neighborhood scale, \textcolor{black}{the patterns of post-flood device-presence change} in Pune and Bandung followed socioeconomic gradients similar to those documented in North American and European disaster research, inviting reconsideration of arguments emphasizing the distinctiveness of disaster response in non-Western settings \cite{bankoff2003cultures, gaillard2010vulnerability, boas_climate_2024}.

\section*{Discussion}

This study investigates post-disaster \textcolor{black}{device-presence change, used here as a proxy for short-term population presence,} in two secondary cities of Monsoon Asia. We combine high-frequency mobile phone data with satellite-derived flood extent polygons to identify which neighborhoods experienced inundation. This approach improves on studies that treat entire cities or metropolitan areas as uniformly "disaster-affected," a common limitation when administrative boundaries rather than hazard footprints define exposure. By mapping flood boundaries onto our mobility grid, we can compare outcomes in flooded versus non-flooded neighborhoods within the same urban region. Examining two cities—Pune and Bandung—with comparable urban functions but different topographic settings and national contexts provides leverage for assessing whether post-flood mobility follows common socioeconomic logics.

Our results reveal a consistent pattern: neighborhoods with lower deprivation scores and stronger economic infrastructure \textcolor{black}{showed increases in detected device presence} following flood events, while more deprived neighborhoods \textcolor{black}{experienced decreases}. This gradient aligns with resource-seeking \textcolor{black}{interpretations} documented in high-income disaster contexts, where \textcolor{black}{increases in detected device presence} tend to concentrate in areas with better services, employment opportunities, and housing quality \cite{hong_measuring_2021, li_mobility_2024}. The consistency of this pattern across Pune and Bandung suggests that economic considerations \textcolor{black}{are associated with} short-term post-disaster mobility in South and Southeast Asian cities in ways that parallel findings from North America and Europe.

The two cities tell a consistent story about deprivation and a more equivocal story about flood exposure itself. In Pune and Bandung, and across every flood-coding specification we examined, neighborhood deprivation was robustly associated with the probability of \textcolor{black}{decreases in detected device presence} (Pune: $\beta_{\text{GRDI}} \approx -0.12$; Bandung: $\beta_{\text{GRDI}} \approx -0.10$; both $p < 0.001$ in the baseline specifications). A one-unit increase in the GRDI lowered the odds of belonging to the \textcolor{black}{device}-increase cluster by roughly 10 to 12 percent in each city, and this gradient did not depend on how we measured inundation at the cell level. By contrast, the main effect of flood exposure was small in both cities and not statistically distinguishable from zero under most codings. Two structural features of the data account for this pattern. Pune's near-uniform inundation footprint left little non-exposed variation to estimate against, and Bandung's flooded and non-flooded cells differed on many correlated dimensions that we cannot fully separate from inundation itself.
 
Whether neighborhood deprivation also amplifies flood exposure was more sensitive to specification. In Pune, the GRDI × flood-overlap interaction was negative and statistically significant under both the $\geq 50\%$ binary coding (Appendix Table A1, Models 4) and the continuous coding (Table~\ref{tab:logit_combined}, Model~2). The implied GRDI slope at the extremes of the overlap distribution (zero versus 100 percent) shifts from near zero in non-overlapping cells to roughly -0.130 in fully overlapping cells, a modest steepening relative to its baseline-model magnitude. In Bandung, the interaction was small and not statistically distinguishable from zero under either the $\geq\!50\%$ binary or the continuous coding, and was only marginally significant under the most lenient $\geq\!1\%$ binary cutoff (Appendix Table A2, Model~2; $p \approx 0.08$).

Several factors may \textcolor{black}{help} explain this divergence. Bandung's topography---a basin surrounded by volcanic highlands with poor natural drainage \cite{tarigan_bandung_2016}---may instead distribute flood impacts in ways that track elevation and drainage more closely than neighborhood deprivation, attenuating the moderating role of socioeconomic conditions. In Pune, more heavily flooded cells showed a steeper deprivation gradient, consistent with flood exposure compounding pre-existing infrastructural deficits in deprived areas \cite{ronco2023exploring, cutter2003social}. Because the flood polygon covered 87\% of the average grid cell, this interaction rested on a narrow band of within-city variation and should be interpreted cautiously.

Our approach makes three contributions to disaster mobility research. First, by linking satellite-derived flood extent polygons to neighborhood-level mobility data, we move beyond the common practice of treating entire region or metropolitan areas as uniformly disaster-affected. Second, our clustering approach, \textcolor{black}{supported by} silhouette analysis and dendrogram inspection, allows mobility patterns to emerge from the data rather than imposing researcher-defined categories. Third, our comparative design—two cities with comparable urban functions \textcolor{black}{and equivalent data quality—strengthens interpretation by enabling cross-city comparison. We also acknowledge important cross-city differences in event timing (July 2019 monsoon flooding in Pune versus the January 2020 New Year flood in Bandung), topography, holiday context, and flood-footprint coverage (approximately 90\% of Pune grid cells fell within the mapped flood polygon).}

Our findings contribute to ongoing discussions about how well disaster mobility frameworks developed in high-income settings travel to other contexts. A body of scholarship has drawn attention to the importance of traditional knowledge, informal housing arrangements, and community-based coping strategies in shaping how populations respond to environmental hazards in the Global South \cite{bankoff2003cultures, whyte_indigenous_2019, adger2013cultural}. Our results complicate this narrative: at the neighborhood scale and in the short term, post-flood \textcolor{black}{device-presence change} in Pune and Bandung followed the same socioeconomic gradient documented in U.S. hurricane studies.

The Pune interaction \textcolor{black}{term}, however, points to an important direction for future research. Recent work on "differential vulnerability" has emphasized that socioeconomic conditions interact with hazard exposure to produce disparate outcomes—a pattern that cannot be captured by examining either factor in isolation \cite{ronco2023exploring, cutter2003social}. The emerging literature on dynamic and multi-hazard vulnerability further highlights how pre-existing social conditions can be amplified by direct hazard exposure, producing compounding effects on affected populations \cite{de_ruiter_challenges_2022, zscheischler2018future}. \textcolor{black}{The tentative indication that neighborhood deprivation moderated this association in Pune but not in Bandung} suggests that the interaction between hazard exposure and neighborhood deprivation deserves systematic attention. Understanding under what local conditions---topography, drainage infrastructure, housing quality---the \textcolor{black}{impact} of flood exposure on \textcolor{black}{detected device presence} strengthens or weakens represents a promising avenue for extending disaster mobility frameworks.

Several limitations warrant discussion. First, the flood exposure measure relies on the Dartmouth Flood Observatory (DFO) archive, which provides approximate flood-affected boundaries compiled from news reports, government sources, and MODIS satellite imagery at roughly 250-meter resolution, rather than precise inundation maps \cite{brakenridge_global_2016, tellman_satellite_2021}. Higher-resolution flood detection products were not available for our study period (2019–2020); the legacy MODIS Water Product that operated from 2011 to April 2022 is no longer accessible, and the current LANCE MCDWD product became available only from 2021 onward. For linking historical flood events to archived mobile phone records, the DFO archive remains the most comprehensive available source offering intra-urban exposure variation, though future research with access to more recent data could leverage improved satellite-based flood detection systems.

Second, mobile phone data capture device movements rather than individual trajectories. The NetMob dataset aggregates unique device counts at the grid-cell level, meaning we observe \textcolor{black}{device-presence change} rather than origin-destination flows. Although the dataset includes origin-destination matrices, privacy protections suppress cells with fewer than ten devices, limiting our ability to trace where \textcolor{black}{increases in detected device presence} originated or to examine individual return \textcolor{black}{trajectories}. Our results should therefore be interpreted as characterizing neighborhood-level outcomes rather than individual evacuation decisions or longer-term residential sorting.

Third, app-based geolocation data likely underrepresent populations without consistent smartphone access—including older adults, low-income individuals, and women—introducing potential sampling bias \cite{zhang_netmob2024_2024}. If the most vulnerable residents are underrepresented in the data, our estimates may understate the true magnitude of deprivation-related \textcolor{black}{decreases in detected device presence}. We cannot directly assess this bias given the scarcity of survey-based disaster mobility research in these South and Southeast Asian contexts; indeed, part of the motivation for this study is that traditional data sources are often unavailable or delayed in rapidly urbanizing Global South cities.

Fourth, data on shelter locations, road network disruptions, and critical infrastructure are scarce for our study contexts. These omissions limit our ability to isolate the causal effect of flood exposure from correlated infrastructure failures. For example, \textcolor{black}{decreases in detected device presence} in flooded areas may reflect not only inundation but also road closures, power outages, or the spatial distribution of emergency shelters—factors we cannot fully disentangle with available data. Future work integrating transportation network data and shelter registries could strengthen causal identification and clarify the mechanisms linking flood exposure to mobility outcomes. 

\textcolor{black}{Fifth, moderate positive spatial autocorrelation in model residuals (Moran's I = 0.187 for Pune, 0.209 for Bandung; Appendix Table A3) suggests that conventional standard errors may be underestimated. The core GRDI coefficients are estimated at $p < 0.001$ in both cities and would remain statistically distinguishable from zero under a two- to threefold increase in standard errors. The interaction evidence is less robust: in the preferred continuous-overlap specification, the GRDI $\times$ flood-overlap interaction reaches conventional thresholds in Pune but not in Bandung, and we treat it as suggestive rather than definitive. The consistency of the GRDI gradient across both cities and its stability across alternative flood-exposure codings (Appendix Tables A1–A2) indicate that the central finding is not an artifact of spatial dependence.}

\textcolor{black}{The timing of the two flood events introduces an additional confound. The Bandung flood began on January 1, 2020, coinciding with the New Year holiday period, when baseline mobility patterns differ from typical weekdays. The Pune flood occurred in early July 2019, outside major Indian public holidays but during monsoon season, when commuting patterns shift seasonally. We cannot fully separate flood-related device-presence changes from these concurrent effects. Because the clustering algorithm partitions neighborhoods relative to one another within the same city and the same time window, however, city-wide shifts from holiday travel or seasonal commuting would affect all neighborhoods similarly and would not by themselves produce the systematic socioeconomic gradient we observe across clusters. A further caveat concerns the temporal alignment of GRDI. Five of its six components are slow-moving structural features whose values were effectively predetermined relative to the 2019 Pune flood; only the 2020 nighttime-lights endpoint postdates the event. To address this, we re-estimated the main specifications using the first principal component of the five pre-event GRDI inputs in place of the published index. The deprivation gradient remained negative and highly significant in both cities (Pune: $\beta = -0.830$, $p < 0.001$; Bandung: $\beta = -0.924$, $p < 0.001$; Appendix Tables~A4--A5), corroborating that the gradient is not driven by the post-event nighttime-lights endpoint.}

\textcolor{black}{Our findings carry practical implications for disaster planning in rapidly urbanizing cities, though each rests on association-based evidence and is best read as a planning consideration or a hypothesis for local testing rather than as a causal prescription.} The consistent socioeconomic gradient in post-flood mobility—with deprived neighborhoods \textcolor{black}{experiencing decreases in detected device presence}—suggests that emergency resource allocation should anticipate differential needs across neighborhoods, targeting support toward areas likely to experience \textcolor{black}{net decreases in detected device presence} that disrupt local services. This recommendation is based on observed associations and should be corroborated by causal studies before large-scale allocation decisions are made. \textcolor{black}{The suggestive Pune interaction raises a further hypothesis for local testing:} that flood mitigation investments—drainage improvements, housing reinforcement, early warning systems—\textcolor{black}{could be especially consequential in more deprived neighborhoods}. Finally, real-time mobility monitoring using aggregated mobile phone data offers a practical tool for humanitarian response, enabling rapid identification of neighborhoods experiencing \textcolor{black}{decreases in detected device presence} to inform shelter placement and resource distribution, while privacy protections appropriately limit access to individual-level information.

\section*{Methods}

\subsubsection*{Data}
\paragraph{Flood Event Spatial Data.} To determine the geographic extent of flooding in our study areas, we obtained flood event polygons from the Dartmouth Flood Observatory (DFO) Global Active Archive of Large Flood Events \cite{brakenridge_global_2016}. The DFO archive, which began in 1993 at Dartmouth College and is now maintained by the Institute of Arctic and Alpine Research (INSTAAR) at the University of Colorado Boulder, provides a complete global record of major flood events from 1985 to the present. This database has been widely used in flood research and epidemiological studies examining the associations between flood exposure and health outcomes \cite{tellman_satellite_2021, najibi_recent_2018}.

The DFO collects data on flood events from multiple sources, including news reports, information from government agencies, Instrumental measurements, and remote-sensing data, all of which are critical for capturing flooding events \cite{brakenridge_global_2016}. Flood events are validated using satellite data, primarily from NASA's Moderate Resolution Imaging Spectroradiometer (MODIS) sensors on the Terra and Aqua satellites. These satellites provide daily global coverage at a 250-meter resolution \cite{brakenridge_modis_2006}. The water detection algorithm developed by DFO uses a ratio of surface-water pixels identified from reflectance values in MODIS Bands 1 and 2, with thresholds applied to Band 7. This method required multiple observations to confirm the presence of water and reduce false positives caused by cloud shadows \cite{brakenridge_flood_2018}. The archived event data includes geographic coordinates (longitude and latitude), time periods (start and end dates), and the affected area. The analysis considers the area in square kilometers, the severity classification (Class 1 for large events occurring every 10–20 years; Class 1.5 for very large events with 20–100 year recurrence intervals; Class 2 for extreme events exceeding 100-year recurrence intervals), the main cause of flooding (such as heavy rain, monsoonal rain, or tropical cyclones), and estimates of human impact, including fatalities and the number of people displaced.

The DFO archive provides GIS polygon files representing the geographic regions affected by each flood event. These files show the geographic areas affected by each flood event. These polygons are created by combining data from news reports and government sources to roughly show the areas affected. They represent the borders of regions impacted by flooding, rather than the exact areas of water \cite{brakenridge_global_2016}. We verified our selection of flood events by cross-referencing it with the EM-DAT International Disaster Database \cite{delforge2023dat}. This was done to confirm the accuracy of the timing and location of significant flood events in Pune, India, and Bandung, Indonesia, and to ensure consistency between the two independent disaster catalogs. 

The DFO archive offers several advantages for this study. It provides uniform global coverage spanning more than 40 years, enabling comparative analysis across different geographical areas using a standardized event classification system. Satellite-based validation using MODIS imagery supplies an objective observational record that complements ground-based reporting. The inclusion of multiple source types — news media, government reports, and remote sensing data — supports triangulation and reduces reliance on any single information channel. The archive is also updated continuously, incorporating information on recent disasters within days of their occurrence. Prior validation studies have confirmed that the DFO archive captures large flood events reliably at global scale, making it appropriate for systematic cross-national comparisons \cite{najibi_recent_2018, tellman_satellite_2021}.

\paragraph{Mobile Phone Geolocation Data.} We analyzed smartphone geolocation data from the NetMob dataset~\cite{zhang_netmob2024_2024}, which contains \textcolor{black}{detected device presence} and origin-destination matrices aggregated at Geohash 5 (approximately 4.9km × 4.9km grid cells) levels for multiple countries, including Indonesia and India. The dataset was developed in collaboration with Cuebiq (also referred to as Spectus) and comprises privacy-preserving aggregated data derived from mobile application location data. These data were collected from users who voluntarily provided informed consent for anonymous data collection for research purposes. The NetMob dataset contains two primary components relevant to our study:

\begin{table}[h]
\centering
\begin{tabular}{|p{3cm}|p{4.5cm}|p{3cm}|p{3cm}|}
\hline
\textbf{Data Type} & \textbf{Description} & \textbf{Spatial Resolution} & \textbf{Temporal Resolution} \\
\hline
\textcolor{black}{Detected Device Presence (DDP)} & Count of unique devices and total observations in geographic areas & Geohash 3 (156km $\times$ 156km) or Geohash 5 (4.9km $\times$ 4.9km) & Daily, 3-hourly \\
\hline
Origin-Destination Matrix (OD) & Flow of users between locations with trip counts and characteristics (duration, length) & Geohash 3 or 5 & Daily, 3-hourly \\
\hline
\end{tabular}
\end{table}

A sample of the \textcolor{black}{Detected Device Presence} data is provided in an example table below\cite{zhang_netmob2024_2024}.

\begin{table}[h]
\centering
\begin{tabular}{|c|c|c|}
\hline
\textbf{geohash\_5} & \textbf{local\_date} & \textbf{no\_of\_unique\_users} \\
\hline
qqg7g & 20190115 & 85 \\
\hline
t9rn6 & 20190116 & 10 \\
\hline
6rfyf & 20190117 & 88 \\
\hline
9u8dq & 20190118 & 95 \\
\hline
\end{tabular}
\end{table}

\textcolor{black}{We used changes in mobile device density at the neighborhood level as a proxy for short-term changes in aggregate device presence around disaster events, not as a direct measure of individual migration or displacement.} For each Geohash 5 cell, we tracked the daily number of unique users before, during, and after the flood events, which allowed us to quantify \textcolor{black}{device-presence} change patterns and recovery trajectories at a neighborhood level.

The data processing methodology employed by the dataset creators ensures privacy protection through several measures, including spatial encoding, temporal aggregation, and minimum threshold requirements (only cells with 10 or more users were included). The original coordinates are transformed to preserve personal privacy, and points falling under sensitive points of interest are removed from the dataset \cite{zhang_netmob2024_2024}. For our analysis, we focused on 2019 data for Pune, India and 2020 data for Bandung, Indonesia, aligning with flood events recorded in these regions. The Bandung flood occurred in January 2020, approximately two months before COVID-19 restrictions took effect in Indonesia, minimizing potential confounding from pandemic-related mobility changes.

\paragraph{Community Socioeconomic Data.} To analyze the relationship between socioeconomic factors and disaster mobility patterns, we incorporated the Global Gridded Relative Deprivation Index (GRDI) Version 1~\cite{ciesin2022grdi}, developed by the Center for International Earth Science Information Network (CIESIN) at Columbia University. The GRDI characterizes relative levels of multidimensional deprivation and poverty at a high spatial resolution of 30 arc-seconds (approximately 1 km), where values range from 0 (lowest deprivation) to 100 (highest deprivation). This index is constructed from six main components: 1. Ratio of Built-up Area to Non-built-up Area (BUILT); 2. Child Dependency Ratio (CDR); 3. Infant Mortality Rates (IMR); 4. Subnational Human Development Index (SHDI); 5. VIIRS Nighttime Lights 2020 (VNL-2020); 6. VIIRS Nighttime Lights Slope 2012-2020 (VNL-slope).

For our analysis, we merged the GRDI data with our Geohash 5 grid cells. Our primary models use the composite GRDI score, which integrates multiple dimensions of deprivation into a single measure; we also show supplementary analyses using a deprivation index based on a PCA score of different sub-components in the Appendix.

\subsubsection*{Measures}
\paragraph{Neighborhood Activity Change.} Following Hong et al.'s\cite{hong_measuring_2021} methodology, we defined daily neighborhood activity levels based on the number of unique devices detected in each Geohash 5 cell. For each neighborhood, we established a baseline activity level by calculating the average number of devices present during a one-month pre-flood period. We then measured the percentage change in daily activity relative to this baseline throughout the study period: $D_{Nd_n} = \frac{U_{Nd_n} - U_{NA}}{U_{NA}}$, where $D_{Nd_n}$ represents the percentage change in activity on day n, $U_{Nd_n}$ is the number of devices in neighborhood N on day n, and $U_{NA}$ is the average number of devices in neighborhood N during the pre-flood baseline period. To reduce noise due to daily fluctuations and highlight underlying trends, we applied a 3-day moving average to the time series data. This smoothing technique helped to eliminate noise while preserving the essential movement patterns related to the disaster events.

\paragraph{Community Deprivation.} To examine the relationship between socioeconomic factors and mobility patterns, we included specific GRDI components in our analysis, particularly focusing on infrastructure quality (BUILT), economic activity (VNL-2020), and economic growth trends (VNL-slope). We also incorporated elevation data \textcolor{black}{from the ASTER Global Digital Elevation Model Version 3~\cite{nasa2019astgtm}} (including both linear and quadratic terms) to account for topographical influences on flood vulnerability and evacuation patterns.
\textcolor{black}{We treat GRDI as capturing slow-moving structural features of neighborhood deprivation that are approximately time-invariant for the analytic window, rather than as a strictly pre-event predictor. Five of its six components (built-up area share, child dependency ratio, infant mortality rate, SHDI, and the VIIRS lights slope) move on multi-year timescales and were effectively predetermined relative to the 2019 Pune event; only the 2020 VIIRS endpoint postdates that event. We report a robustness check using the first principal component of the five pre-event inputs in Appendix Table A4-A5.}

\subsubsection*{Models and Analytic Approach}
\paragraph{Neighborhood Clustering Analysis.} To identify distinct patterns in disaster response and recovery, we implemented an agglomerative hierarchical clustering algorithm on the neighborhood activity time series data. The clustering was performed on the smoothed time series of percentage change in activity levels, covering the period from 30 days before through 30 days after each flood onset.

Similar to Hong et al. (2021), we used Ward's method to calculate the distance between clusters, which minimizes the variance within clusters while maximizing the variance between different groups. The mathematical formulation for Ward's distance between clusters $C_i$ and $C_j$ is:
\begin{equation}
D_w(C_i, C_j) = \sum_{x \in C_i}(x - r_i)^2 + \sum_{x \in C_j}(x - r_j)^2 - \sum_{x \in C_{ij}}(x - r_{ij})^2,    
\end{equation}
where $C_{ij}$ represents the merged cluster of $C_i$ and $C_j$, and $r_i$, $r_j$, and $r_{ij}$ are the respective cluster centroids. We recognize that data-driven metrics tend to favor a two-cluster solution across both locations. Silhouette scores, which measure cluster cohesion and separation, peaked at k=2 for both Pune (0.35) and Bandung (0.30), declining steadily with additional clusters (Appendix Figure A1). The elbow method similarly indicated diminishing returns in within-cluster variance reduction beyond two clusters. The hierarchical dendrograms further support this choice: both cities show a clear bifurcation into two major branches at the highest merge level (height $\approx 7.5$ for Pune; $\approx 10.5$ for Bandung), with subsequent splits occurring at substantially lower heights. These two clusters correspond to neighborhoods that \textcolor{black}{experienced increases versus decreases in detected device presence} during flood events—a substantively meaningful distinction that captures the primary dimension of post-disaster mobility response. Descriptive labels \textcolor{black}{("increase in detected device presence" and "decrease in detected device presence")} were assigned post hoc based on observed mean trajectories within each cluster.

\paragraph{Binomial Logistic Regression.} We estimate binomial logistic regression models to examine how neighborhood characteristics predict cluster membership. The dependent variable is a dichotomous indicator distinguishing neighborhoods that experienced \textcolor{black}{increases in detected device presence} from those that experienced \textcolor{black}{decreases in detected device presence}, as determined by the hierarchical clustering analysis described above. Independent variables include the Global Relative Deprivation Index (GRDI), \textcolor{black}{a continuous indicator of flood exposure (the percentage of cell area overlapping the DFO flood polygon)}, elevation (linear and quadratic terms), and an interaction between GRDI and flood exposure. The model can be expressed as: 
\begin{equation}
\log\left(\frac{P(Y = 1)}{1 - P(Y = 1)}\right) = \alpha + \beta X,    
\end{equation}

\textcolor{black}{where $Y = 1$ indicates membership in the cluster with increases in detected device presence, $\alpha$ is the intercept, and $\beta$ is a vector of coefficients. The vector of predictor variables, $X$, comprises the composite GRDI score, flood-exposure overlap (the continuous percentage of each Geohash-5 cell's area that intersects the DFO flood polygon, 0–100), and elevation entered as both linear and quadratic terms.} This modeling approach allowed us to identify the socioeconomic, geographical, and flood exposure factors associated with different neighborhood mobility patterns during flood events while controlling for potential confounding variables, without imposing assumptions about the ordering of response categories.

\textcolor{black}{Flood exposure is measured as the percentage of each Geohash-5 cell's area that intersects the Dartmouth Flood Observatory (DFO) flood-event polygon, ranging from 0 to 100. We adopt this continuous coding for two reasons. First, the underlying inundation signal is itself continuous, and partial overlap is common at the cell boundaries of an approximately $4.9\,\text{km} \times 4.9\,\text{km}$ grid; collapsing this variation into a single yes/no indicator discards information that is informative for cells along the flood margin. Second, a continuous measure avoids the arbitrariness of a threshold choice, which is particularly consequential in Pune, where roughly 90\% of analytic cells fall inside the DFO polygon and any lenient binary cutoff would classify nearly the entire city as flood-exposed. To verify that our central findings do not depend on this measurement choice, we re-estimated each city's model under three alternative codings: a binary indicator with a $\geq\!1\%$ overlap threshold (the most inclusive coding), a binary indicator with a $\geq\!50\%$ overlap threshold, and the continuous specification used in the main text. Appendix Tables A1 and A2 report the alternative specifications with and without the GRDI $\times$ flood-exposure interaction. The deprivation gradient on GRDI is statistically significant and substantively comparable across every coding in both cities (Pune $\approx -0.12$; Bandung $\approx -0.10$; both $p<0.001$ in the baseline columns), \textcolor{black}{indicating} that the relationship between neighborhood deprivation and post-flood \textcolor{black}{device-presence} change is not an artifact of how flood exposure is coded at the cell level. The flood-exposure main effect itself is small and not statistically distinguishable from zero in Pune under any coding and negative but imprecisely estimated in Bandung; we therefore interpret the flood-exposure coefficient descriptively rather than causally throughout the manuscript.}
 
We excluded grid cells intersected by major railway corridors to reduce confounding from commuter flows unrelated to flood-associated \textcolor{black}{decreases in detected device presence}. Railway stations and transit hubs can generate large fluctuations in device counts that reflect daily commuting patterns rather than residential mobility. Removing these cells yields a more conservative estimate of flood-related \textcolor{black}{device-presence change}. After applying this restriction and dropping cells with more than three consecutive days of missing mobility data, the analytic sample includes 159 grid cells for Pune and 301 grid cells for Bandung.

\section*{Data Availability}

\textcolor{black}{The mobile-device data are the NetMob 2024 Data Challenge dataset
\cite{zhang_netmob2024_2024}, comprising daily unique-device counts and
origin-destination matrices at Geohash-5 resolution. The dataset is stored in
the World Bank Development Data Catalog under accession 0066094
(\url{https://datacatalog.worldbank.org/search/dataset/0066094/aggregated_mobility_and_density_data_for_the_netmob_2024_data_challenge}).
Access is controlled because the underlying location data are personally
sensitive: the World Bank grants access on request under a custom licence and
the authors may not redistribute the data. Requests are submitted through the
Data Request Form on the catalogue page, with enquiries to
\url{data@worldbank.org} and a response typically within four weeks. The
governing terms limit recipients to non-commercial research and prohibit
re-identification and onward redistribution.}

\textcolor{black}{The remaining data are openly available from their providers: flood extent
polygons for events 4775 and 4859 from the Dartmouth Flood Observatory Global
Active Archive of Large Flood Events \cite{brakenridge_global_2016},
\url{https://web.archive.org/web/20210526085436/http://floodobservatory.colorado.edu/Archives/}; the Global Gridded
Relative Deprivation Index Version 1 \cite{ciesin2022grdi},
\url{https://doi.org/10.7927/3XXE-AP97}; and the ASTER Global Digital Elevation
Model Version 3~\cite{nasa2019astgtm},
\url{https://doi.org/10.5067/ASTER/ASTGTM.003}.}

\textcolor{black}{The derived data underlying all reported results are deposited with the
analysis code and reproduce every table and figure without access to the
restricted source data. Further data are available on reasonable request from
the corresponding author.}

\section*{Code Availability}

\textcolor{black}{The Python and Stata code that reproduces all tables and figures is available
at \url{https://github.com/turtleman99/post_flood_mobility} under the
MIT Licence, together with the derived datasets needed to re-run the clustering
and the regression models. Analyses used Python 3.12 and Stata 16.}

\begin{samepage}
\section*{Acknowledgements}
The authors acknowledge the contributions of Cuebiq, the World Bank, and the NetMob 2024 organizing committee in enabling access to the datasets analyzed in this study. Earlier versions of this paper were presented at NetMob 2024, the 2025 Population Association of America Annual Meeting, the 2025 American Sociological Association Mid-Year Methodology Section Conference, and the 2025 American Sociological Association Annual Meeting. The authors are grateful to the discussants and participants at these meetings for their valuable comments.\par
\end{samepage}

\section*{Funding}
This research received no funding.

\section*{Author Contributions}
\noindent Conceptualization: SXY, GW, SC, ZC, CC\\
Data curation: SXY, GW\\
Formal analysis: SXY, GW\\
Investigation: SXY, GW\\
Methodology: SXY, GW\\
Project administration: CC\\
Supervision: CC\\
Validation: SXY, GW\\
Visualization: SXY, GW\\
Writing -- original draft: SXY, GW, SC, CC\\
Writing -- review \& editing: SXY, GW, SC, ZC, CC

\section*{Competing Interests}
The authors declare no competing interests.

\bibliography{references}

\newpage
\section*{Tables and Figures}

\noindent\begin{minipage}{\textwidth}
\small 
\textit{Notes:} Base map: Esri, TomTom, Garmin, FAO, NOAA, USGS, © OpenStreetMap contributors, and the GIS User Community. The figure shows the geographical context of our two study areas. Panels (a) and (b) depict Pune, India: (a) shows Pune's location within India, highlighted in red, and (b) provides a detailed view of the Pune district with city center marked by a red dot. Mumbai, a major metropolitan area with over 10 million residents, is indicated by a larger red dot to the northwest. Panels (c) and (d) show Bandung, Indonesia: (c) displays Bandung's location on Java Island within the Indonesian archipelago, highlighted in red, and (d) shows the detailed boundaries of the Bandung Metropolitan Area. The city center is marked with a red dot, while Jakarta, Indonesia's capital and largest city, is indicated by a larger red dot to the northwest. The surrounding topography of Bandung includes mountainous terrain, particularly to the north and east.
\end{minipage}

\newpage
\vspace{0.5cm}
\noindent\begin{minipage}{\textwidth}
\small 
\textit{Notes:} The two groups (decrease and increase) are based on the agglomerative hierarchical clustering algorithm applied to mobility patterns observed before, during, and after flooding events. Analysis was conducted using a grid system with 4.9 km-by-4.9 km cells (Geohash level 5). \textcolor{black}{(a) Normalized deviation in device presence (y-axis) relative to baseline, by day relative to flood onset (x-axis, day 0 = onset; shaded band = event window), for the two clusters identified in Pune: a Decrease cluster (teal, n = 66 neighborhoods) and an Increase cluster (orange, n = 93 neighborhoods); \textcolor{black}{semi-transparent thin lines show individual grid trajectories, and shaded ribbons show $\pm 1$ s.d. across neighborhoods within each cluster.} (b) Map of Pune showing the spatial distribution of grid cells assigned to the Increase (orange) and Decrease (teal) clusters shown in (a). (c) As in (a), for Bandung: Decrease cluster (teal, n = 167 neighborhoods), Increase cluster (orange, n = 134 neighborhoods). (d) Map of Bandung showing the spatial distribution of grid cells assigned to the Increase (orange) and Decrease (teal) clusters shown in (c).}
\end{minipage}

\newpage
\vspace{0.01cm}
\noindent\begin{minipage}{0.98\textwidth}
\small 
\textit{Notes:} The two groups are based on the agglomerative hierarchical clustering algorithm applied to mobility patterns observed before, during, and after flooding events. Analysis was conducted using a grid system with 4.9km-by-4.9km cells. The mobility patterns were derived from smartphone geolocation data from the NetMob 2024 dataset, which contains \textcolor{black}{detected device presence} and origin-destination matrices provided by Cuebiq (Spectus). Box plots show the distribution of the Gridded Relative Deprivation Index (GRDI) across grid cells in the Decrease and Increase mobility-response groups. (a) Pune: left, map of grid cells colored by mobility-response cluster assignment from Figure 2b (orange = Increase, teal = Decrease); right, box plot of the Gridded Relative Deprivation Index (GRDI) for grid cells in the Decrease (n = 66) and Increase (n = 93) clusters. Box shows median and interquartile range; whiskers extend to 1.5× the interquartile range; open circles are outliers. (b) As in (a), for Bandung: Decrease (n = 167) and Increase (n = 134) clusters.
\end{minipage}

\newpage
\begin{table}[ht]
\centering
\small
\caption{Neighborhood Cluster Characteristics in Pune, India, and Bandung, Indonesia}
\begin{tabular}{lcc|cc|c}
\hline
\hline
\multicolumn{6}{l}{flood ID = 4775, Pune, India (\textit{N}=159)} \\
\hline
 & \multicolumn{2}{c|}{Overall} & \textcolor{black}{Device-Presence} & \textcolor{black}{Device-Presence} & \\
Variable & \multicolumn{2}{c|}{} & Decrease & Increase & Sig.(F) \\
\cline{2-5}
 & Mean & SD & Mean & Mean & \\
\hline
\textcolor{black}{Grid overlap with flood (\%)} & \textcolor{black}{87.354} & \textcolor{black}{32.480} & \textcolor{black}{80.123} & \textcolor{black}{92.485} & \textcolor{black}{*} \\
Relative Deprivation Index & 39.671 & 13.395 & 48.483 & 33.417 & *** \\
Built-up Infrastructure & 55.831 & 23.588 & 41.798 & 65.790 & *** \\
Nighttime Lights & 52.172 & 27.540 & 33.653 & 65.315 & *** \\
Nighttime Lights change overtime & 64.298 & 19.819 & 52.764 & 72.484 & *** \\
Elevation (10 m) & 59.916 & 8.626 & 61.641 & 58.691 & * \\
\hline
\hline
\\[5pt]
\hline
\hline
\multicolumn{6}{l}{flood ID = 4859, Bandung, Indonesia (\textit{N}=301)} \\
\hline
 & \multicolumn{2}{c|}{Overall} & \textcolor{black}{Device-Presence} & \textcolor{black}{Device-Presence} & \\
Variable & \multicolumn{2}{c|}{} & Decrease & Increase & Sig.(F) \\
\cline{2-5}
 & Mean & SD & Mean & Mean & \\
\hline
\textcolor{black}{Grid overlap with flood (\%)} & \textcolor{black}{68.504} & \textcolor{black}{45.228} & \textcolor{black}{64.689} & \textcolor{black}{73.257} &  \textcolor{black}{-} \\
Relative Deprivation Index & 32.715 & 13.367 & 39.360 & 24.432 & *** \\
Built-up Infrastructure & 17.647 & 14.285 & 12.262 & 20.377 & *** \\
Nighttime Lights & 50.648 & 25.734 & 34.953 & 59.630 & *** \\
Nighttime Lights change overtime & 38.412 & 25.050 & 24.311 & 46.483 & *** \\
Elevation (10 m) & 33.568 & 34.561 & 42.469 & 22.475 & *** \\
\hline
\hline
\end{tabular}
\\[7pt]
\begin{tabular}{p{0.76\linewidth}}
\small \textit{Notes:} $^{*}p<0.05$, $^{**}p<0.01$, $^{***}p<0.001$. \textcolor{black}{Significance in the Sig.(F) column is based on one-way analysis-of-variance (ANOVA) omnibus F-tests comparing means between the Decrease and Increase clusters for each variable.} The two groups \textcolor{black}{(decrease and increase in detected device presence)} are based on the agglomerative hierarchical clustering algorithm applied to mobility patterns observed before, during, and after flooding events. Flood exposure is measured as the percentage of each Geohash-5 grid cell's area that intersects the Dartmouth Flood Observatory (DFO) flood-event polygon, ranging from 0 (no overlap) to 100 (full overlap). This continuous coding preserves the underlying spatial variation in inundation extent and avoids the arbitrariness of a binary threshold; alternative binary specifications and their robustness implications are reported in Appendix Tables A1--A2. The Global Gridded Relative Deprivation Index (GRDI) is a composite measure of multidimensional deprivation, with higher values indicating greater deprivation (i.e., poorer socioeconomic conditions). The GRDI integrates six component indicators: Built-up Infrastructure, Child Dependency Ratio, Infant Mortality Rate, Subnational Human Development Index (SHDI), Nighttime Lights (VNL 2020), and Nighttime Lights change over time (VNL slope 2012--2020). For each component, higher raw values indicate more favorable conditions (e.g., greater infrastructure density, higher human development, stronger economic activity), except for Child Dependency Ratio and Infant Mortality Rate, where higher values indicate less favorable conditions. Components indicating favorable conditions are reverse-coded prior to aggregation, ensuring that all contribute positively to the final deprivation score. GRDI data were obtained from the Center for International Earth Science Information Network (CIESIN), Columbia University (Version 1; DOI: https://doi.org/10.7927/3xxe-ap97). Because of minimal variance across grid cells within the study area, Child Dependency Ratio, Infant Mortality Rate, and SHDI were excluded from the regression models.
\end{tabular}
\label{cluster_characteristics}
\end{table}

\newpage
\begin{table}[htbp]
\centering
\small
\caption{\textcolor{black}{Binomial Logistic Regressions Predicting \textcolor{black}{Device-Presence} Change}}
\label{tab:logit_combined}
\begin{tabular}{lcccc}
\toprule
 & \multicolumn{2}{c}{Pune (Disaster 4775)} & \multicolumn{2}{c}{Bandung (Disaster 4859)} \\
\cmidrule(lr){2-3} \cmidrule(lr){4-5}
 & (1) Baseline & (2) {+}Interaction & (3) Baseline & (4) {+}Interaction \\
\midrule
GRDI (avg)                & -0.116*** & -0.004  & -0.104*** & -0.072**  \\
                          & (0.021)   & (0.066) & (0.013)   & (0.025)   \\
Flood overlap (\%)        & 0.010     & 0.069*  & -0.007    & 0.009     \\
                          & (0.006)   & (0.034) & (0.004)   & (0.011)   \\
GRDI $\times$ Flood overlap &           & -0.0013$^{\dagger}$ &           & -0.0001    \\
                          &           & (0.0007) &           & (0.0001)   \\
Elevation (10 m) & -0.061 & -0.068 & -0.013  & -0.008   \\
                          & (0.108)   & (0.114) & (0.016) & (0.017)  \\
Elevation (10 m) squared         & 0.000    & 0.000  & -0.0001  & -0.0001   \\
                          & (0.001)   & (0.001) & (0.0001) & (0.0001)  \\
Constant                  & 7.825*     & 2.686  & 3.988***& 2.744**  \\
                          & (3.389)   & (4.515) & (0.639) & (1.026)  \\
\midrule
Observations              & 159       & 159     & 301     & 301      \\
Pseudo $R^2$              & 0.296     & 0.310   & 0.273    & 0.278    \\
\bottomrule
\end{tabular}
\\[7pt]
\noindent\begin{minipage}{0.7\textwidth}
\small
\textit{Notes:} Standard errors in parentheses. $^{\dagger}\,p<0.1$, $^{*}\,p<0.05$, $^{**}\,p<0.01$, $^{***}\,p<0.001$. Significance stars are based on two-sided Wald $z$-tests on each coefficient. Dependent variable: binary indicator equal to 1 if the cell belongs to the \textcolor{black}{cluster with increases in detected device presence} and 0 if it belongs to the \textcolor{black}{cluster with decreases in detected device presence}, as determined by the hierarchical clustering analysis. Flood exposure is measured as the continuous percentage of each Geohash-5 cell's area that intersects the Dartmouth Flood Observatory polygon (0--100). Columns (2) and (4) add the GRDI $\times$ flood-overlap interaction. Alternative binary codings ($\geq\!1\%$ and $\geq\!50\%$ overlap) are reported in Appendix Tables~A1 (Pune) and~A2 (Bandung).
\end{minipage}
\end{table}

\singlespacing
\newpage
\section*{Appendix}

% Reset numbering and prefix with "A" for the appendix
\setcounter{table}{0}
\setcounter{figure}{0}
\renewcommand{\thetable}{A\arabic{table}}
\renewcommand{\thefigure}{A\arabic{figure}}

\begin{figure}[htbp]
    \centering
    
    % Row 1: Dendrograms
    \begin{subfigure}[b]{0.46\textwidth}
        \centering
        \includegraphics[
            width=\linewidth,
            height=0.22\textheight,
            keepaspectratio]{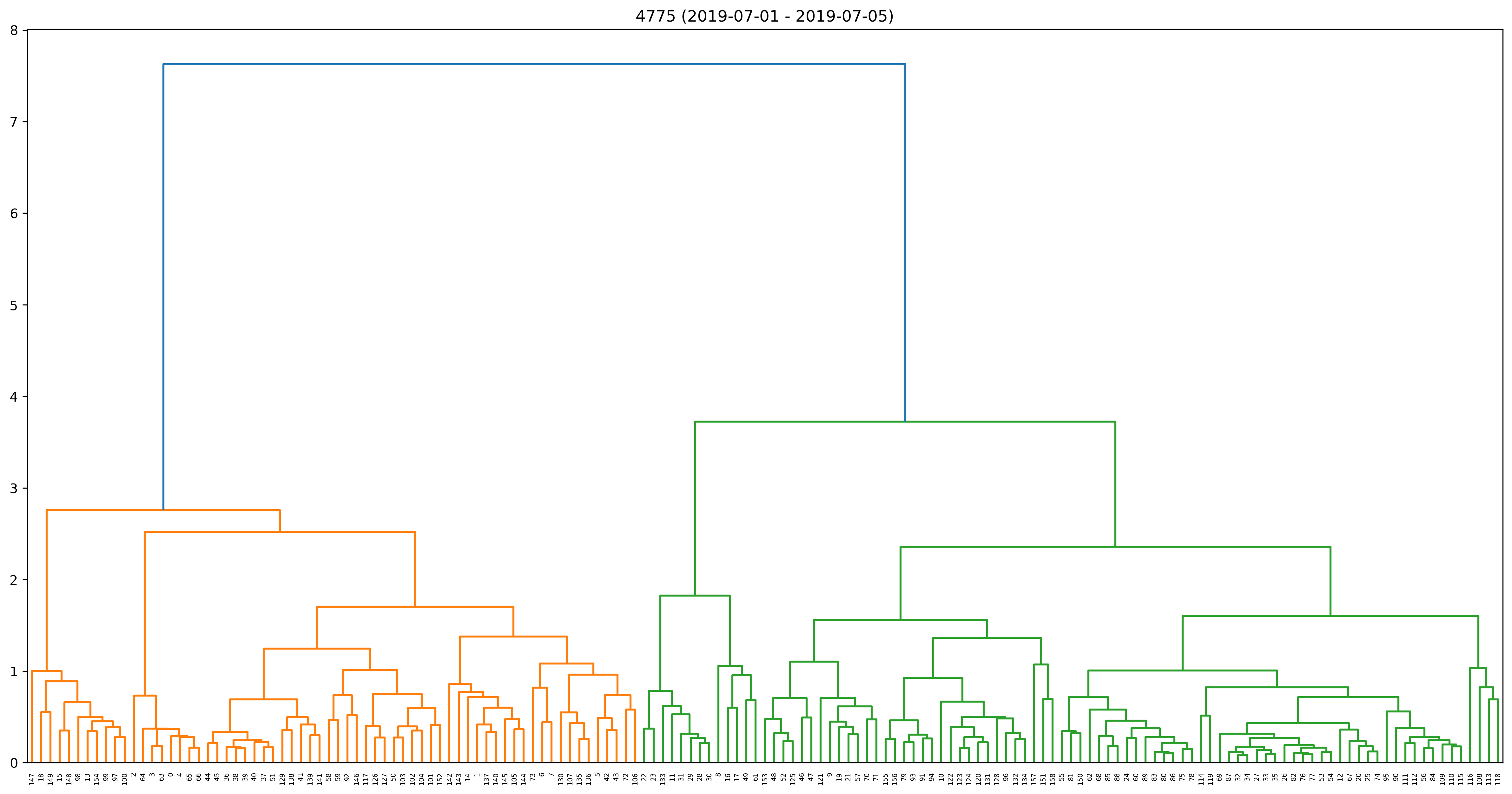}
        \caption{Dendrogram: Pune, India}
        \label{fig:dendrogram_pune}
    \end{subfigure}
    \hfill
    \begin{subfigure}[b]{0.46\textwidth}
        \centering
        \includegraphics[
            width=\linewidth,
            height=0.22\textheight,
            keepaspectratio]{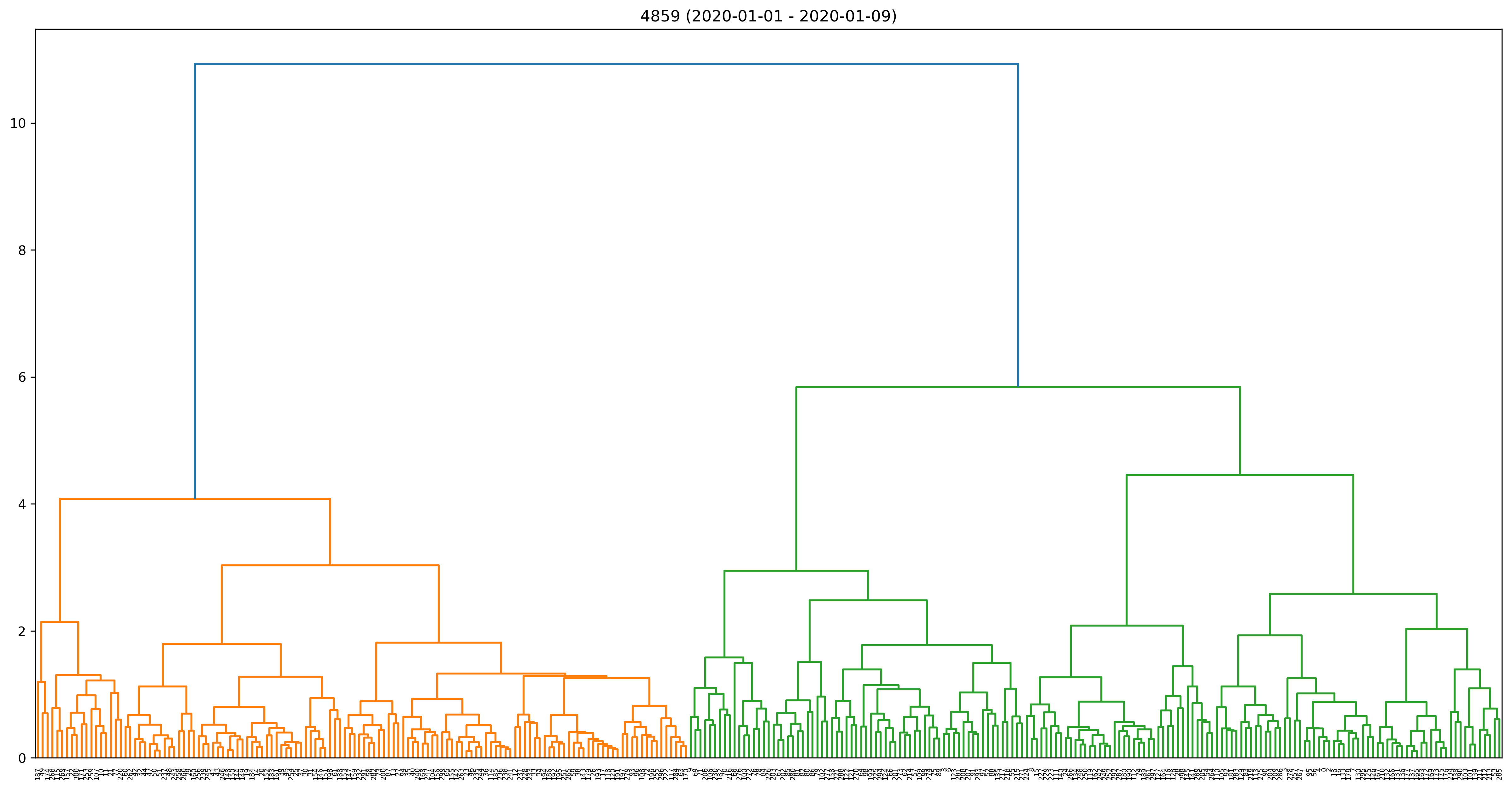}
        \caption{Dendrogram: Bandung, Indonesia}
        \label{fig:dendrogram_bandung}
    \end{subfigure}
    
    \vspace{0.05cm}
    
    % Row 2: Silhouette Analysis
    \begin{subfigure}[b]{0.46\textwidth}
        \centering
        \includegraphics[
            width=\linewidth,
            height=0.22\textheight,
            keepaspectratio]{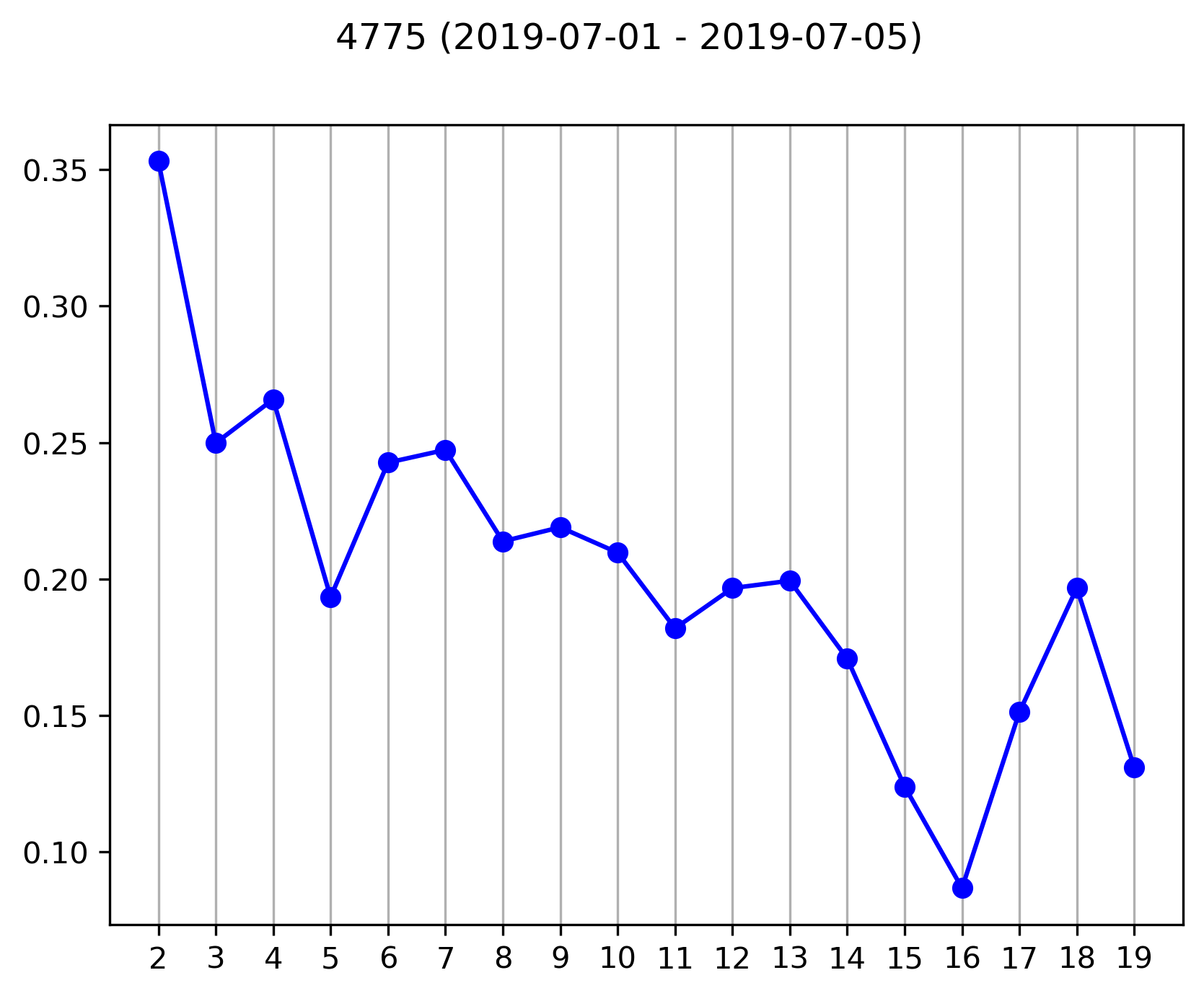}
        \caption{Silhouette Analysis: Pune, India}
        \label{fig:silhouette_pune}
    \end{subfigure}
    \hfill
    \begin{subfigure}[b]{0.46\textwidth}
        \centering
        \includegraphics[
            width=\linewidth,
            height=0.22\textheight,
            keepaspectratio]{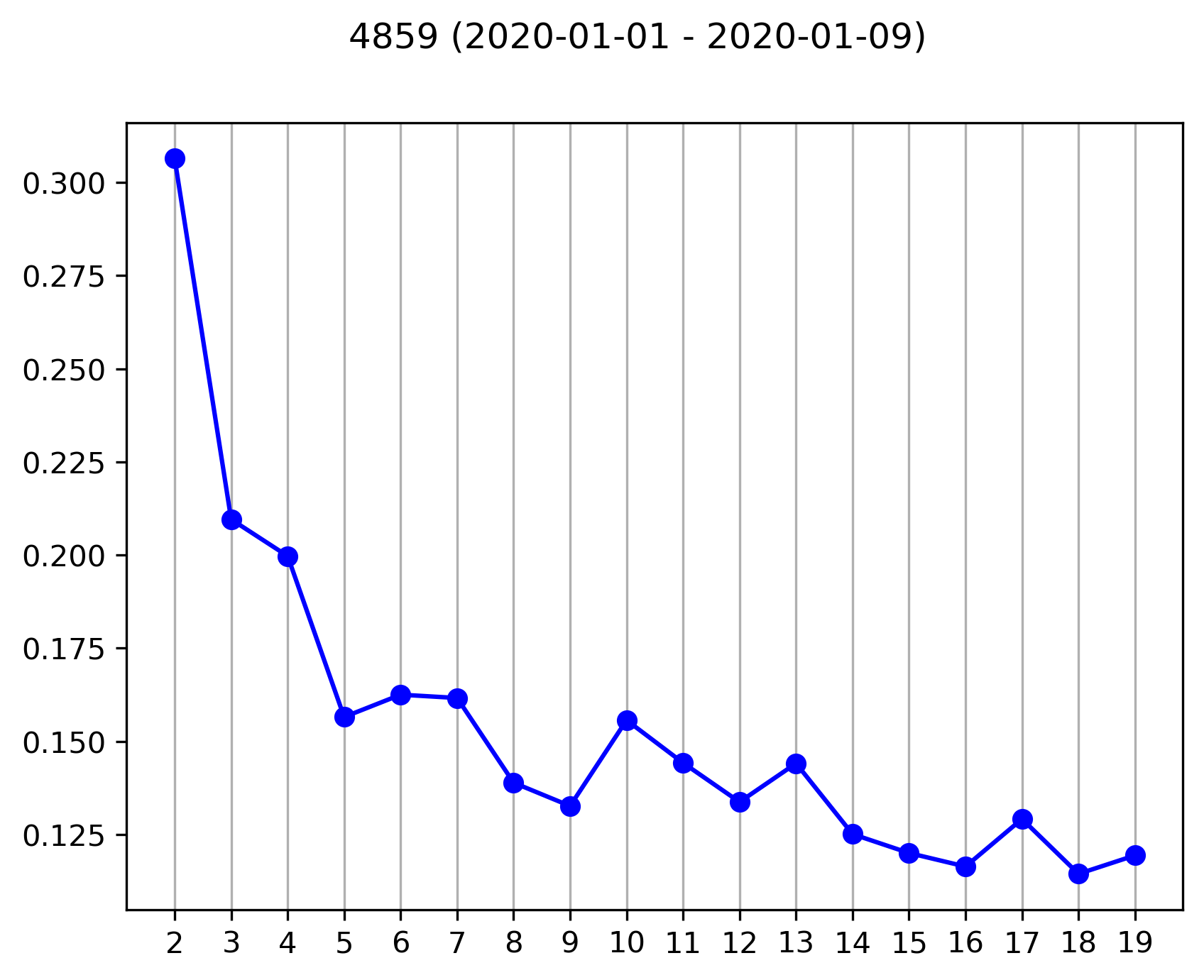}
        \caption{Silhouette Analysis: Bandung, Indonesia}
        \label{fig:silhouette_bandung}
    \end{subfigure}
    
    \vspace{0.05cm}
    
    % Row 3: Elbow Method
    \begin{subfigure}[b]{0.46\textwidth}
        \centering
        \includegraphics[
            width=\linewidth,
            height=0.22\textheight,
            keepaspectratio]{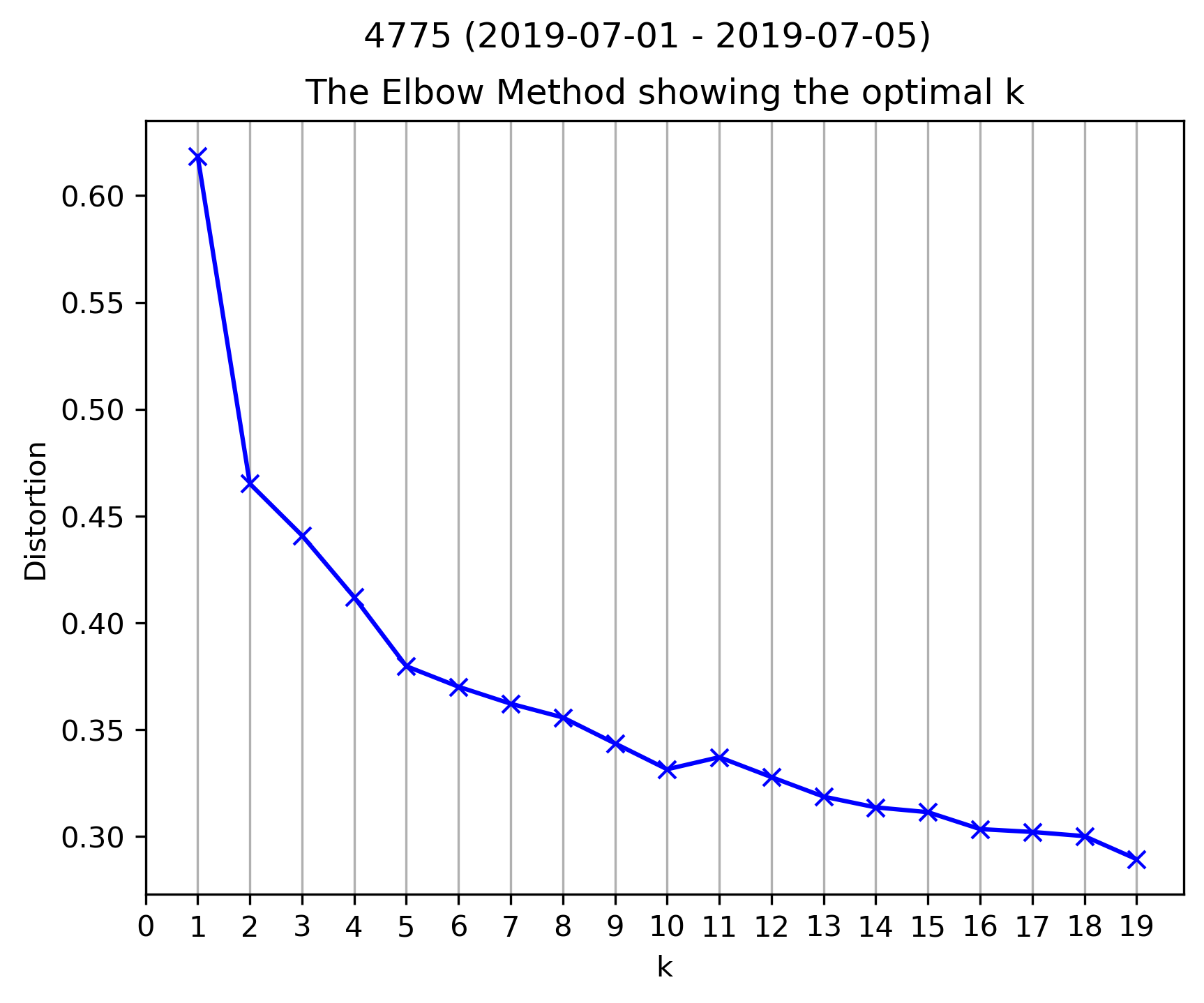}
        \caption{Elbow Method: Pune, India}
        \label{fig:elbow_pune}
    \end{subfigure}
    \hfill
    \begin{subfigure}[b]{0.46\textwidth}
        \centering
        \includegraphics[
            width=\linewidth,
            height=0.22\textheight,
            keepaspectratio]{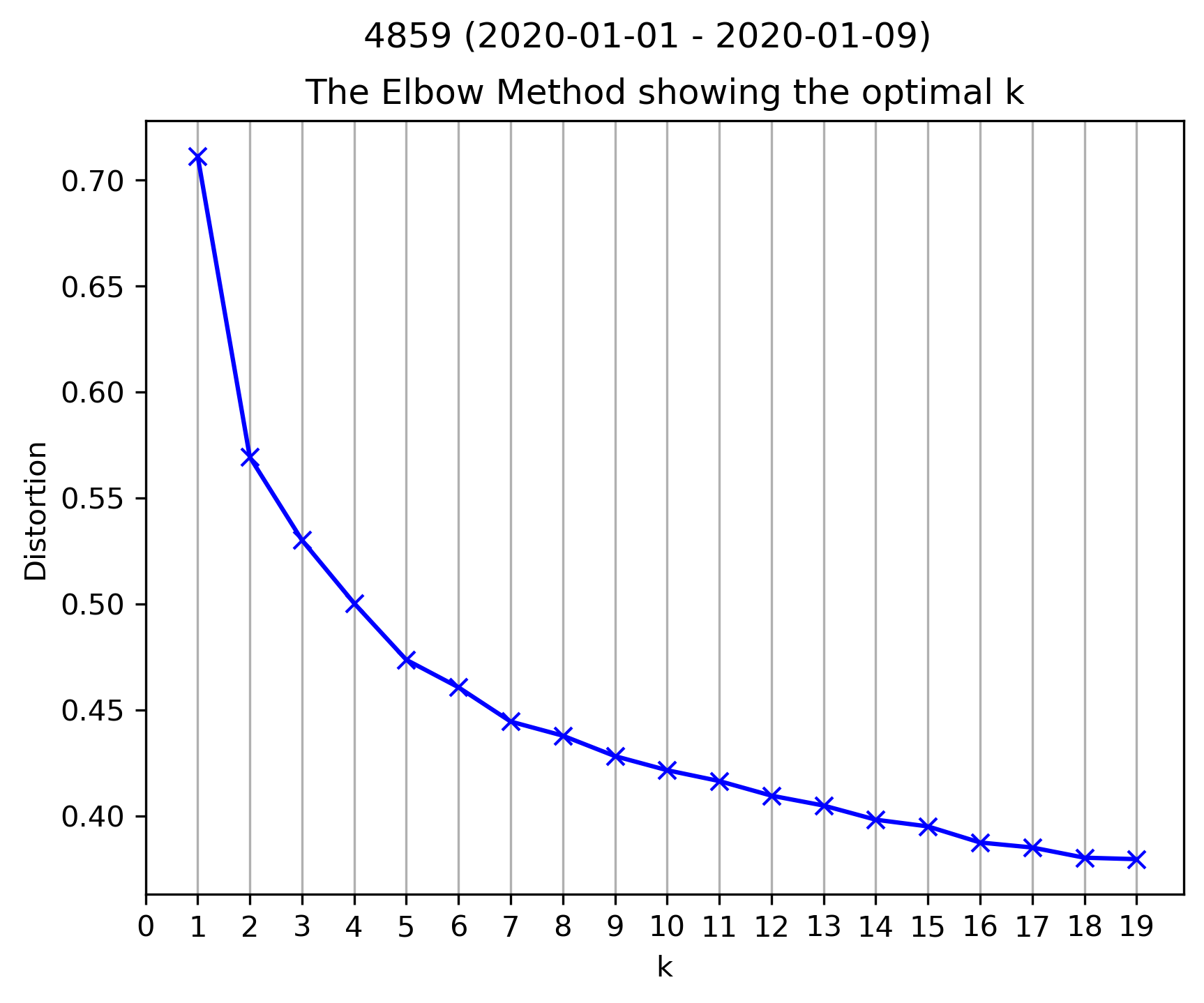}
        \caption{Elbow Method: Bandung, Indonesia}
        \label{fig:elbow_bandung}
    \end{subfigure}
    
    \caption{Cluster Validation Diagnostics for Pune, India, and Bandung, Indonesia}
    \label{fig:elevation_analysis}
\end{figure}

\vspace{0.3cm}
\begin{minipage}{\textwidth}
\small 
\textit{Notes:} This figure presents three complementary diagnostics used to evaluate cluster solutions for the mobility time-series data. Panels (a)--(b) show hierarchical clustering dendrograms, Panels (c)--(d) show silhouette analysis, and Panels (e)--(f) show elbow method results.

\textbf{Dendrograms (Panels a--b):} The dendrograms display Ward's linkage distances at which grid cells merge into larger clusters. For Pune, two primary branches separate at a height of $\approx$ 7.5, with further sub-structure emerging between heights 2.5 and 4. For Bandung, a distinct branch joins the main cluster only at a height $\approx$ 14, with additional subdivisions at a height $\approx$ 7.

\textbf{Silhouette Analysis (Panels c--d):} Silhouette scores provide a measure of how similar observations are to their own cluster compared to other clusters; higher scores indicate better-defined groupings. In Pune, the two-cluster solution gives the highest average silhouette score (0.35). Although $k=4$ (0.27) shows a slight improvement over $k=3$ (0.25), the two-cluster solution is still the most effective. For Bandung, silhouette scores decline from $k=2$ \textcolor{black}{(0.30)} through $k=4$ (0.21).

\textbf{Elbow Method (Panels e--f):} The elbow plots display within-cluster distortion against the number of clusters. For Pune, distortion declines gradually with a subtle inflection around $k=4$--5. For Bandung, the curve shows modest flattening beginning around $k=3$--4. Neither plot exhibits a sharp elbow.

\end{minipage}

% ── Table 1: Pune ────────────────────────────────────────────────────────────
\begin{table}[htbp]
\centering
\small
\caption{\textcolor{black}{Robustness of the deprivation gradient to alternative binary flood-exposure codings, Pune (Disaster 4775)}}
\label{tab:appendix-pune-binary}
\begin{tabular}{lcccc}
\toprule
 & \multicolumn{2}{c}{Binary, $\geq\!1\%$ overlap} & \multicolumn{2}{c}{Binary, $\geq\!50\%$ overlap} \\
\cmidrule(lr){2-3} \cmidrule(lr){4-5}
 & (1) Baseline & (2) {+}Interaction & (3) Baseline & (4) {+}Interaction \\
\midrule
GRDI (avg)                  & -0.116*** & -0.053  & -0.092*** & 0.026   \\
                            & (0.022)   & (0.041) & (0.020)   & (0.053) \\
Flooded (binary)            & 0.887     & 3.012   & 0.309     & 6.679*  \\
                            & (0.637)   & (2.108) & (0.573)   & (2.770) \\
GRDI $\times$ Flooded  &           & -0.066  &           & -0.135* \\
                            &           & (0.052) &           & (0.058) \\
Elevation (10 m) & 0.005     & 0.002   & 0.006     & 0.000   \\
                            & (0.114)   & (0.115) & (0.114)   & (0.118) \\
Elevation (10 m) squared           & -0.001    & -0.001  & -0.001    & -0.001  \\
                            & (0.001)   & (0.001) & (0.001)   & (0.001) \\
Constant                    & 6.541     & 4.118   & 6.743     & 1.270   \\
                            & (3.522)   & (3.747) & (3.550)   & (4.295) \\
\midrule
Observations                & 159       & 159     & 159       & 159     \\
Log-likelihood              & -77.31    & -76.41  & -77.89    & -75.20  \\
Pseudo $R^2$                & 0.246     & 0.255   & 0.240     & 0.267   \\
\bottomrule
\end{tabular}
\begin{flushleft}
\footnotesize Standard errors in parentheses. $^{*}\,p<0.05$, $^{**}\,p<0.01$, $^{***}\,p<0.001$. Significance stars are based on two-sided Wald $z$-tests on each coefficient. Columns (1)--(2) define the binary flood indicator as one for any DFO-polygon overlap ($\geq\!1\%$); columns (3)--(4) restrict this indicator to cells with $\geq\!50\%$ polygon overlap. Columns (2) and (4) add the GRDI $\times$ flood-exposure interaction. The continuous-overlap specification reported in main-text Table~2 is the preferred coding because of the high share of analytic cells inside the DFO polygon in Pune (approximately 90\%) and the arbitrariness of any single binary threshold.
\end{flushleft}
\end{table}

% ── Table 2: Bandung ─────────────────────────────────────────────────────────
\begin{table}[htbp]
\centering
\small
\caption{\textcolor{black}{Robustness of the deprivation gradient to alternative binary flood-exposure codings, Bandung (Disaster 4859)}}
\label{tab:appendix-bandung-binary}
\begin{tabular}{lcccc}
\toprule
 & \multicolumn{2}{c}{Binary, $\geq\!1\%$ overlap} & \multicolumn{2}{c}{Binary, $\geq\!50\%$ overlap} \\
\cmidrule(lr){2-3} \cmidrule(lr){4-5}
 & (1) Baseline & (2) {+}Interaction & (3) Baseline & (4) {+}Interaction \\
\midrule
GRDI (avg)                  & -0.101*** & -0.056  & -0.104*** & -0.077** \\
                            & (0.014)   & (0.029) & (0.013)   & (0.024)  \\
Flooded (binary)            & 0.608     & 2.553*  & -0.675    & 0.638    \\
                            & (0.317)   & (1.171) & (0.372)   & (1.067)  \\
GRDI $\times$ Flooded &           & -0.058  &           & -0.038   \\
                            &           & (0.030) &           & (0.029)  \\
Elevation (10 m) & -0.013    & -0.011  & -0.012    & -0.008   \\
                            & (0.016)   & (0.016) & (0.016)   & (0.017)  \\
Elevation (10 m) squared           & -0.000    & -0.000  & -0.000    & -0.000   \\
                            & (0.000)   & (0.000) & (0.000)   & (0.000)  \\
Constant                    & 3.711***  & 2.412*  & 3.949***  & 2.910**  \\
                            & (0.616)   & (1.025) & (0.624)   & (0.990)  \\
\midrule
Observations                & 301       & 301     & 301       & 301      \\
Log-likelihood              & -149.92   & -148.34 & -150.405  & -149.585 \\
Pseudo $R^2$                & 0.275     & 0.283   & 0.273     & 0.277    \\
\bottomrule
\end{tabular}
\begin{flushleft}
\footnotesize Standard errors in parentheses. $^{*}\,p<0.05$, $^{**}\,p<0.01$, $^{***}\,p<0.001$. gnificance stars are based on two-sided Wald $z$-tests on each coefficient. Columns (1)--(2) define the binary flood indicator as one for any DFO-polygon overlap ($\geq\!1\%$); columns (3)--(4) restrict it to cells with $\geq\!50\%$ polygon overlap. Columns (2) and (4) add the GRDI $\times$ flood-exposure interaction. The continuous-overlap specification reported in main-text Table~2 is the preferred coding because the GRDI gradient is invariant to coding while the interaction term is not.
\end{flushleft}
\end{table}

\newpage
% ── Table: Moran's I ──
\begin{table}[htbp]
\centering
\small
\caption{\textcolor{black}{Moran's $I$ test for spatial autocorrelation in model residuals}}
\label{tab:morans-i}
\begin{tabular}{lccccc}
\toprule
City & $N$ & Moran's $I$ & $\text{E}[I]$ & $z$ & $p$ (two-sided) \\
\midrule
Pune (Disaster 4775)    & 159 & $+$0.187 & $-$0.006 & $+$5.313 & $<$0.001 \\
Bandung (Disaster 4859) & 301 & $+$0.209 & $-$0.003 & $+$7.841 & $<$0.001 \\
\bottomrule
\end{tabular}
\begin{flushleft}
\footnotesize Moran's $I$ is computed on the Pearson residuals of the baseline continuous-overlap model (column~3 in each city's robustness table) using row-standardized 8-nearest-neighbor weights; inference uses the randomization variance.
\end{flushleft}
\end{table}

\newpage
\begin{table}[!htbp]
\centering
\caption{\textcolor{black}{Principal component analysis on the five pre-event GRDI inputs}}
\label{tab:appendix_pca_5comp}
\small
\renewcommand{\arraystretch}{1.1}
\begin{tabular}{lrrrrr|rrrrr}
\toprule
 & \multicolumn{5}{c}{\textbf{Pune (Disaster 4775)}} & \multicolumn{5}{c}{\textbf{Bandung (Disaster 4859)}} \\
\cmidrule(lr){2-6}\cmidrule(lr){7-11}
Component & PC1 & PC2 & PC3 & PC4 & PC5 & PC1 & PC2 & PC3 & PC4 & PC5 \\
\midrule
Built-up area share & -0.645 & -0.052 & -0.204 & +0.178 & +0.713 & +0.347 & +0.751 & -0.084 & +0.093 & -0.548 \\
Child dependency ratio & +0.211 & +0.728 & -0.327 & +0.565 & +0.009 & +0.384 & -0.471 & +0.612 & +0.209 & -0.461 \\
Infant mortality rate & -0.218 & +0.206 & +0.894 & +0.334 & -0.010 & +0.335 & -0.262 & -0.598 & +0.677 & +0.060 \\
Subnational HDI & +0.295 & -0.648 & -0.040 & +0.700 & +0.034 & -0.410 & +0.320 & +0.440 & +0.695 & +0.231 \\
VIIRS lights slope & -0.636 & -0.078 & -0.226 & +0.217 & -0.700 & +0.672 & +0.207 & +0.259 & -0.081 & +0.656 \\
\midrule
Eigenvalue & 1.961 & 1.056 & 0.970 & 0.782 & 0.230 & 1.423 & 1.156 & 1.017 & 0.863 & 0.540 \\
Proportion of variance & 0.392 & 0.211 & 0.194 & 0.156 & 0.046 & 0.285 & 0.231 & 0.203 & 0.173 & 0.108 \\
Cumulative & 0.392 & 0.603 & 0.798 & 0.954 & 1.000 & 0.285 & 0.516 & 0.719 & 0.892 & 1.000 \\
\midrule
$N$ & \multicolumn{5}{c}{159} & \multicolumn{5}{c}{301} \\
\bottomrule
\end{tabular}
\end{table}

\begin{table}[htbp]
\centering
\small
\caption{\textcolor{black}{Binomial Logistic Regressions Predicting \textcolor{black}{Device-Presence} Change --- Robustness Check Using a 5-Component Pre-Event GRDI (First Principal Component)}}
\label{tab:appendix_logit_5comp}
\begin{tabular}{lcccc}
\toprule
 & \multicolumn{2}{c}{Pune (Disaster 4775)} & \multicolumn{2}{c}{Bandung (Disaster 4859)} \\
\cmidrule(lr){2-3} \cmidrule(lr){4-5}
 & (1) Baseline & (2) {+}Interaction & (3) Baseline & (4) {+}Interaction \\
\midrule
GRDI (PC1, 5-comp.) & -0.830$^{***}$ & -0.373 & -0.924$^{***}$ & -0.940$^{*}$ \\
                          & (0.169) & (0.585) & (0.229) & (0.395) \\
Flood overlap (\%) & 0.011 & 0.014 & -0.014$^{*}$ & -0.013$^{*}$ \\
                          & (0.006) & (0.008) & (0.005) & (0.006) \\
GRDI(PC1) $\times$ Flood overlap &  & -0.005 &  & 0.0001 \\
                          &  & (0.006) &  & (0.005) \\
Elevation (10 m) & -0.073 & -0.056 & -0.020 & -0.021 \\
                          & (0.108) & (0.105) & (0.025) & (0.025) \\
Elevation (10 m) squared & 0.000 & -0.0001 & -0.0001 & -0.0001 \\
                          & (0.001) & (0.001) & (0.0001) & (0.0001) \\
Constant & 3.758 & 2.816 & 0.351 & 0.351 \\
                          & (3.205) & (3.257) & (0.589) & (0.590) \\
\midrule
Observations              & 159       & 159     & 301     & 301      \\
Pseudo $R^2$              & 0.223     & 0.226   & 0.221   & 0.221   \\
\bottomrule
\end{tabular}
\\[7pt]
\begin{minipage}{0.85\textwidth}
\small
\textit{Notes:} Standard errors in parentheses. $^{*}\,p<0.05$, $^{**}\,p<0.01$, $^{***}\,p<0.001$. Significance stars are based on two-sided Wald $z$-tests on each coefficient. Dependent variable: binary indicator equal to 1 if the cell belongs to the \textcolor{black}{cluster with increases in detected device presence} and 0 if it belongs to the \textcolor{black}{cluster with decreases in detected device presence}, as determined by the hierarchical clustering analysis. The deprivation regressor is the first principal component of five pre-event GRDI inputs (built-up area share, child dependency ratio, infant mortality rate, Subnational HDI, and VIIRS lights slope); the 2020 VIIRS endpoint is excluded. PC1 is oriented so that it correlates positively with the published six-component GRDI ($r=+0.93$ in Pune, $r=+0.84$ in Bandung). For Pune, built-up share and VIIRS slope are reverse-coded; Flood exposure is the continuous percentage of each Geohash-5 cell's area that intersects the Dartmouth Flood Observatory polygon (0--100). Columns (2) and (4) add the GRDI(PC1) $\times$ flood-overlap interaction.
\end{minipage}
\end{table}

\newpage
\begin{table}[htbp]
\centering
\small
\caption{\textcolor{black}{Deprivation Gradient Across Alternative Cluster Solutions}}
\label{tab:grdi_krobust}
\begin{tabular}{lcc}
\hline
\hline
Contrast & Neighborhoods ($n$) & GRDI Score \\
 & (contrast vs.\ reference) & \\
\hline
\multicolumn{3}{l}{\textit{Panel A: Pune, India} ($n=159$)} \\
\hline
Two clusters (binary coding, Table A1) & & \\
\quad Increase vs.\ Decrease & 93 vs.\ 66 & $-0.116^{***}$ \\
 & & (0.021) \\
Three clusters & & \\
\quad Slight Increase vs.\ Decrease & 80 vs.\ 66 & $-0.122^{***}$ \\
 & & (0.022) \\
\quad Great Increase vs.\ Decrease & 13 vs.\ 66 & $-0.087^{**}$ \\
 & & (0.031) \\
Four clusters & & \\
\quad Great Increase vs.\ Great Decrease & 13 vs.\ 55 & $-0.078^{*}$ \\
 & & (0.031) \\
\quad Slight Increase vs.\ Great Decrease & 80 vs.\ 55 & $-0.115^{***}$ \\
 & & (0.023) \\
\quad Slight Decrease vs.\ Great Decrease & 11 vs.\ 55 & 0.062 \\
 & & (0.054) \\
\hline
\multicolumn{3}{l}{\textit{Panel B: Bandung, Indonesia} ($n=301$)} \\
\hline
Two clusters (binary coding, Table A2) & & \\
\quad Increase vs.\ Decrease & 134 vs.\ 167 & $-0.101^{***}$ \\
 & & (0.013) \\
Three clusters & & \\
\quad Increase vs.\ Great Decrease & 134 vs.\ 98 & $-0.091^{***}$ \\
 & & (0.014) \\
\quad Stable vs.\ Great Decrease & 69 vs.\ 98 & $0.032^{*}$ \\
 & & (0.015) \\
Four clusters & & \\
\quad Increase vs.\ Great Decrease & 134 vs.\ 59 & $-0.112^{***}$ \\
 & & (0.017) \\
\quad Stable vs.\ Great Decrease & 69 vs.\ 59 & 0.014 \\
 & & (0.017) \\
\quad Slight Decrease vs.\ Great Decrease & 39 vs.\ 59 & $-0.044^{*}$ \\
 & & (0.018) \\
\hline
\hline
\end{tabular}
\\[7pt]
\begin{minipage}{0.9\textwidth}
\small
\textit{Notes:} $^{*}p<0.05$, $^{**}p<0.01$, $^{***}p<0.001$. Standard errors in parentheses. Each entry is the deprivation coefficient from one outcome equation of a multinomial logistic regression estimated with the same covariates as the main models (binomial in the two-cluster case); within each city and solution, the most declining cluster is the reference category. The middle column gives the number of neighborhoods in the contrast cluster and in the reference cluster; within each solution, the cluster sizes sum to the panel total. Cluster names describe each cluster's mean change in detected device presence 30 days after flood onset within its own solution; clusters carrying the same name at different numbers of clusters are not necessarily the same set of neighborhoods, and the counts identify them. In every solution in both cities, each cluster with rising detected device presence is significantly less deprived than the most declining cluster, with magnitudes close to the two-cluster estimates. Contrasts among declining and stable clusters are smaller and less consistently signed, as expected, since they do not compare rising with declining trajectories.
\end{minipage}
\end{table}

\end{document}